\documentclass[lettersize,journal]{IEEEtran}
\usepackage{amsmath,amsfonts}
\usepackage{algorithmic}
\usepackage{algorithm}
\usepackage{array}
\usepackage[caption=false,font=normalsize,labelfont=sf,textfont=sf]{subfig}
\usepackage{textcomp}
\usepackage{stfloats}
\usepackage{url}
\usepackage{verbatim}
\usepackage{graphicx}
\usepackage{cite}

\usepackage{multirow}
\usepackage{multicol}
\usepackage{booktabs}
\usepackage[section]{placeins}
\usepackage{color}
\usepackage{threeparttable}

\newcommand{\Rmnum}[1]{\expandafter\@slowromancap\romannumeral #1@}
\usepackage{xcolor}

\usepackage{makecell}

\usepackage{threeparttable}
\usepackage{diagbox}
\usepackage{hyperref}

\begin{document}

\title{Exploring Transferability of Multimodal Adversarial Samples for Vision-Language Pre-training Models with Contrastive Learning}

% \title{Boosting Multimodal Adversarial Attacks with Contrastive Learning for Vision-Language Pre-training Models   }

\author{Youze Wang, Wenbo Hu, Yinpeng Dong, Hanwang Zhang, Hang Su, Richang Hong,~\IEEEmembership{Member,~IEEE,}
        % <-this % stops a space

\thanks{Manuscript received July 22, 2024; revised November 9, 2024; accepted January 6, 2025.
This work is jointly supported by National Natural Science Foundation of China (No. U23B2031 and 62306098), the Open Projects Program of State Key Laboratory of Multimodal Artificial Intelligence Systems, Funds for the Central Universities (No. JZ2024HGTB0256), the SMP-IDATA Open Youth Fund (No. SMP-2023-iData-009),
the Open Project of Anhui Provincial Key Laboratory of Multimodal Cognitive Computation, Anhui University (No. MMC202412) and the National Research Foundation, Singapore under its AI Singapore Programme (AISG Award No: AISG2-RP-2021-022).}
\thanks{Y. Wang, W. Hu and R. Hong are at the School of Computer Science and Information Engineering, Hefei University of Technology, Hefei 230009, China. (e-mail: $\{$wenbohu, hongrc$\}$@hfut.edu.cn) }
\thanks{Y. Dong and H. Su are at Tsinghua University, Beijing, 100084, China. (e-mail: {dongyinpeng@mail.tsinghua.edu.cn, suhangss@tsinghua.edu.cn )}}
\thanks{H. Zhang is with the School of Computer Science and Engineering, Nanyang Technological University, Singapore 639798.
 (e-mail: hanwangzhang@gmail.com)}
 % \thanks{This work is jointly supported by National Natural Science Foundation of China (No. 62306098), the Open Projects Program of State Key Laboratory of Multimodal Artificial Intelligence Systems, the Fundamental Research Funds for the Central Universities (No. JZ2024HGTB0256) and the National Research Foundation, Singapore under its AI Singapore Programme (AISG Award No: AISG2-RP-2021-022).}
 \thanks{Corresponding author: Wenbo Hu.}
 }
% \thanks{
% © © 20xx IEEE. Personal use of this material is permitted. Permission from IEEE must be obtained for all other uses, in any current or future media, including reprinting/republishing this material for advertising or promotional purposes, creating new collective works, for resale or redistribution to servers or lists, or reuse of any copyrighted component of this work in other works.

% This work was supported by the National Key Research and Development Program of China (No. 2022ZD0118802), the NSF of China Project (No. 61932009 and 62306098), and the Open Projects Program of State Key Laboratory of Multimodal Artificial Intelligence Systems. (Corresponding author: Wenbo Hu).

% Y. Wang, W. Hu, R. Hong are at the School of Computer Science and Information Engineering, Hefei University of Technology, Hefei 230009, China. (E-mail: wangyouze@mail.hfut.edu.cn, \{wenbohu,hongrc\}@hfut.edu.cn) }}

        % <-this % stops a space

\markboth{IEEE Transactions on Multimedia}%
{Shell \MakeLowercase{\textit{et al.}}: A Sample Article Using IEEEtran.cls for IEEE Journals}


\maketitle

\begin{abstract}
 The integration of visual and textual data in Vision-Language Pre-training (VLP) models is crucial for enhancing vision-language understanding. However, the adversarial robustness of these models, especially in the alignment of image-text features, has not yet been sufficiently explored. In this paper, we introduce a novel gradient-based multimodal adversarial attack method, underpinned by contrastive learning, to improve the transferability of multimodal adversarial samples in VLP models. This method concurrently generates adversarial texts and images within imperceptive perturbation, employing both image-text and intra-modal contrastive loss. We evaluate the effectiveness of our approach on image-text retrieval and visual entailment tasks, using publicly available datasets in a black-box setting. Extensive experiments indicate a significant advancement over existing single-modal transfer-based adversarial attack methods and current multimodal adversarial attack approaches.
\end{abstract}

\begin{IEEEkeywords}
Vision-and-Language Pre-training, Multimodal, Adversarial Attack, Adversarial Transferability.
\end{IEEEkeywords}
 
\section{Introduction}
\IEEEPARstart{V}{ision}-Language Pre-training (VLP) models are pivotal in integrating visual and textual data, which is considered as a significant research focus aimed at enhancing vision-language understanding~\cite{li2021align, yang2022vision, li2022blip, li2023blip}. Contrastive learning aligns cross-modal features effectively by attracting matched image-text pairs and repelling mismatched ones, demonstrating its capability to learn transferable multimodal features without labeled data~\cite{radford2021learning}.
This alignment strategy in VLP models has led to substantial advancements in applications such as image-text retrieval~\cite{cao2022image}, visual entailment~\cite{xie2019visual}, and visual question answering~\cite{haohccl}, highlighting their transformative potential in multimodal understanding.
\begin{figure}[ht]
\centering\includegraphics[width=3.6in, height=2in]{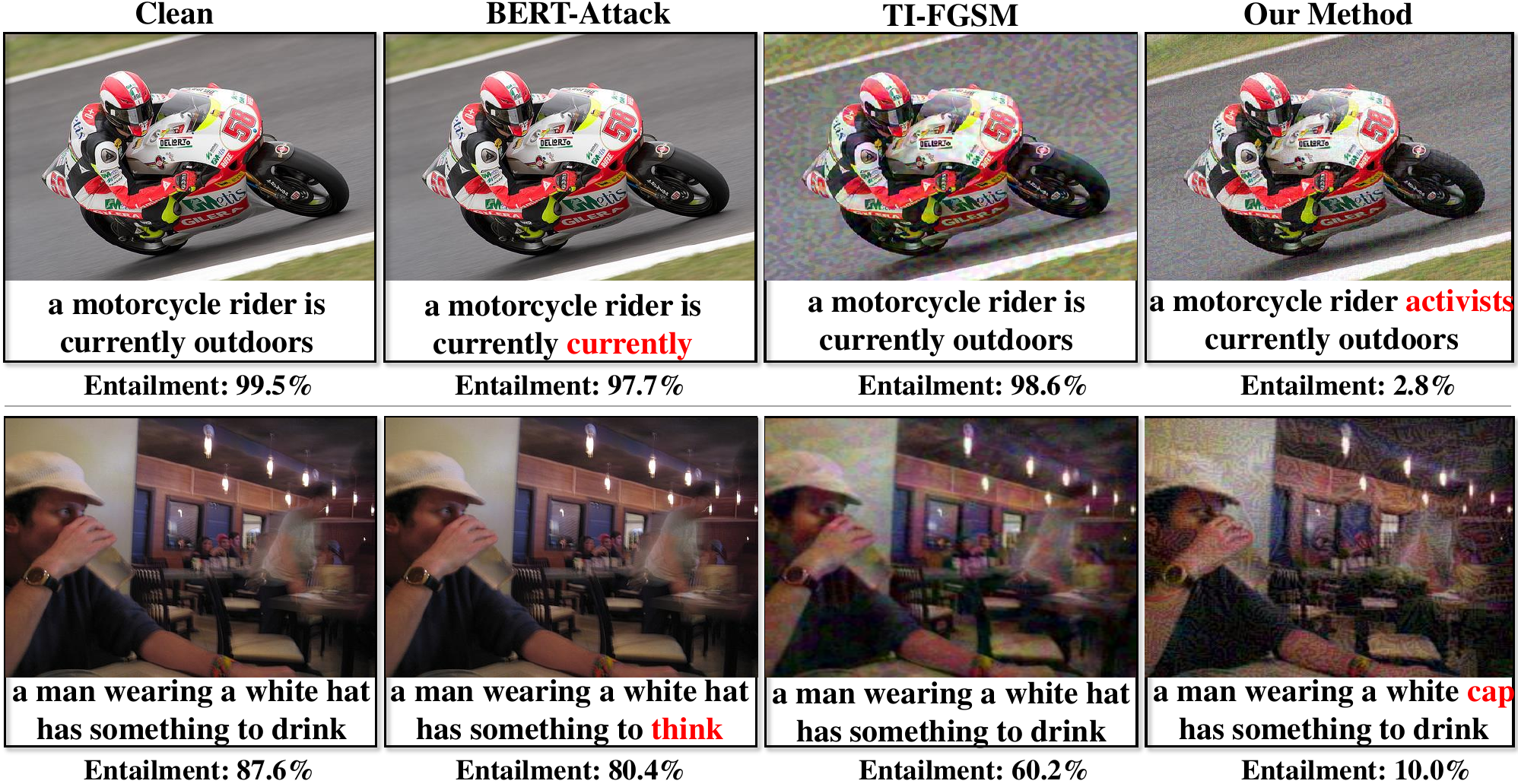}
\caption{In a black-box setting, we show three adversarial samples generated by our methods and single-modal adversarial attack methods (TI-FGSM~\cite{dong2019evading} and Bert-Attack~\cite{li2020bert}). 
\textbf{Top row}: the victim model is TCL~\cite{yang2022vision}. \textbf{Bottom row}: the victim model is ALBEF~\cite{li2021align}. The probabilities of these adversarial samples by the victim models are also provided below each pair of adversarial samples.}
 \label{fig:intr}
\end{figure}

It is well studied that the adversarial robustness would bring vulnerability by adding the minor input adversarial perturbations, either textual~\cite{li2018textbugger,li2020bert},  or visual~\cite{goodfellow2014explaining, madry2017towards}.
Based on it, transfer-based adversarial attacks are proposed which are considered more practical in real-world applications~\cite{dong2019evading, xie2019improving,guo2020backpropagating,naseer2021generating}.
Some preliminary efforts have been undertaken to explore the adversarial robustness of the VLP models. 
The Co-attack method proposed to give a sequential text-then-image approach to generate multi-modal adversarial samples in a white-box setting~\cite{zhang2022towards}. 
The SGA approach extends the Co-attack framework by leveraging cross-modal interactions across multiple alignments with set-level guidance, enhancing the transferability of multimodal adversarial samples in the black-box setting~\cite{lu2023set}.
The robustness of the large VLP models is evaluated under the task of image-text generation and considers the transferability in a box-box style. Though comprehensive, this evaluation study is limited to the adversarial images~\cite{zhao2023evaluating}.

It is worth noting that currently adversarial strategies in VLP models often overlook the complex interplay between text and image modalities. Effective multimodal adversarial attacks must target each modality while considering the contextual relation of the adversarial pair, which enhances their efficacy in VLP models, as depicted in Figure~\ref{fig:intr}. Given the discrete nature of texts and the continuous nature of images, developing a unified framework for generating transferable adversarial text-image pairs remains a critical and unexplored area.

In this paper, we propose a gradient-based multimodal adversarial attack method with contrastive learning (VLP-attack) to generate transferable multimodal adversarial samples against VLP models. Our approach unifies textual and visual adversarial attacks within a gradient-based framework, enabling simultaneous production of adversarial texts and images under perceptibility constraints. We employ image-text and intra-modal contrastive losses to enhance sample transferability. Our method is validated using CLIP as the surrogate model, outperforming existing single-modal and multimodal adversarial attack methods in image-text retrieval and visual entailment tasks. Victim models tested include the CLIP series~\cite{ilharco_gabriel_2021_5143773, sun2023eva}, ALBEF~\cite{li2021align}, TCL~\cite{yang2022vision}, BLIP~\cite{li2022blip}, BLIP2~\cite{li2023blip}, and MiniGPT-4~\cite{zhu2023minigpt}.

In summary, the main contributions are listed as follows: 
 \begin{itemize}
    \item Diverging from previous methods, our research employs gradient-based techniques to simultaneously optimize adversarial text and images, investigating transfer-based multimodal adversarial attacks against VLP models.
     \item To effectively disturb the alignment of image-text representation in a black-box setting, we provide contrastive learning with sufficient image-text variations to perturb the inherent structures in the benign samples and the contextual integrity of image-text pairs from different views.
    \item We conduct comprehensive experiments and analysis to demonstrate our method outperforms other baselines including single-modal and multimodal adversarial attacks.
 \end{itemize}

% -------------------------------------------------------------------------------------------------------------------------------
 \section{Related Works}
 \label{sec:related_work}
 In this section, we introduce the related works, including visional language pretraining, self-supervised contrastive learning, and black box adversarial attack and its transferability.
\subsection{Vision-Language Pre-training (VLP) and Contrastive Learning}
Recent advancements in VLP models have been substantial. These models, trained on extensive datasets of images and corresponding textual descriptions, aim to synthesize understanding between visual and textual data, enhancing vision-language tasks. One such groundbreaking model is CLIP~\cite{radford2021learning}, which excels in image-text similarity and zero-shot multimodal tasks,
and demonstrates a sophisticated capability to concurrently process and reason about both visual and textual information. 
ALBEF~\cite{li2021align}, aligning image and text representations before fusing them with a multimodal encoder, further enhances its learning capabilities from noisy web data through a momentum distillation process. 
% To counteract the issue of similar inputs from the same modality diverging due to training on noisy data, 
TCL~\cite{yang2022vision} employs triple contrastive learning to harness complementary information from different modalities, thus improving representation learning and alignment. 
BLIP~\cite{li2022blip} leverages a dataset bootstrapped from large-scale noisy image-text pairs to pre-train a multimodal encoder-decoder model.
% which is adept for both understanding-based and generation-based tasks. 
Building upon this, BLIP2~\cite{li2023blip} introduces a lightweight Querying Transformer to achieve effective vision-language alignment with frozen unimodal models.
MiniGPT-4~\cite{zhu2023minigpt} marks another stride in this field, aligning the visual model with a large language model, making it particularly suitable for image-grounded text generation tasks. 

Contrastive learning~\cite{radford2021learning}, known for its excellent alignment of text-image features for VLP models, encourages the learned features away from the invariant of augmentations of negative samples, resulting in more generalizable and intrinsic attributes across different models.  To enrich the feature space, data augmentation techniques~\cite{wu2020clear, meng2021coco, bayer2022survey}  usually be employed to generate augmentations for image-text pairs.
Collectively, these VLP models with contrastive learning underscore the importance of aligning image and text features to enhance performance on various downstream tasks.
However, existing transfer-based attacks, primarily focused on single-modal data~\cite{dong2019evading, lin2019nesterov, zhao2023evaluating, wang2021enhancing, wang2021feature}, have been inadequate for more complex tasks requiring a nuanced understanding of multimodal information.

%\wenbo{this subsection seems to be overlapped with sec 2.1. What about this: 2.1-VLP, 2.2-adversial attack and transferability, 2.3-multimodal adversial attack? please note that remember to mention the challenge we want to solve at the very end}
\subsection{Black-box Adversarial Attack and Transferability }
Black-box adversarial attacks and their transferability have gained significant attention in recent years due to the vulnerability of machine-learning models to adversarial samples~\cite{wang2021feature, zhu2022toward, sadrizadeh2023transfool, dong2021query}. Compared with the white-box adversarial attack~\cite{goodfellow2014explaining, wang2023iterative},  the black-box attack assuming no knowledge about the target model is a harder setting to study adversarial transferability. 
In the transfer-based black-box setting,  the text adversarial attack methods: Bert-attack~\cite{li2020bert}, PWWS~\cite{jin2020bert},  GBDA~\cite{guo2021gradient},  and image adversarial attack methods: MI-FGSM~\cite{dong2018boosting}, TI-FGSM~\cite{dong2019evading}, SI-NI-FGSM~\cite{lin2019nesterov}, C-GSP~\cite{yang2022boosting} and~\cite{li2021adversarial, gao2021push}, generate adversarial samples on surrogate models at a white-box setting to attack other target models, even when the attacker has no access to the target model's architecture, gradients, or parameters. 
Among them, LPM~\cite{wei2023boosting} posits that model-specific discriminative regions contribute significantly to overfitting on the source model. To address this, it employs a differential evolutionary algorithm to identify a patch-wise binary mask. This mask is used to drop out image patches associated with these model-specific regions, thereby enhancing the transferability of adversarial examples.
However, given the complexity of multimodal data, traditional transfer-based strategies against vision-only models, designed for single-modal data, fall short in effectively perturbing the alignment between text and image features in VLP models. Crafting multimodal adversarial samples with high transferability thus presents a significant challenge.
% In response, our work leverages the principles of contrastive learning~~\cite{chen2020simple,gao2021simcse,radford2021learning} to generate a diverse array of image-text variations. These variations play a crucial role in disrupting the underlying structures of benign samples from multiple angles. This approach not only facilitates the creation of highly transferable multimodal adversarial samples but also enhances our understanding of the vulnerabilities inherent in VLP models. 

% To fully evaluate the adversarial robustness of VLP models in a black-box setting,
% we need to craft multimodal adversarial samples with high transferability. However, to our knowledge, attacking the images and texts against VLP modes in black-box settings remains unexplored.
%$\wenbo{please refine here and 2.3}
\subsection{Multimodal Adversarial Attack}
While Vision-Language Pretraining (VLP) models have achieved remarkable advancements in various vision-language tasks, the adversarial robustness of these models, particularly the alignment of image and text features, remains unexplored. Zhang's work~\cite{zhang2022towards} delves into adversarial attacks on VLP models in a white-box setting, offering valuable insights for designing multimodal adversarial attacks and enhancing model robustness. Similarly, Wang~\cite{wang2023iterative} introduces a white-box adversarial attack specifically targeting multimodal text generation tasks. 
Zhao~\cite{zhao2023evaluating} investigates the robustness of large VLP models through perturbing the images modality in a black-box setting.
Lu~\cite{lu2023set} enhances the transferability of multimodal adversarial samples by utilizing cross-modal interactions and data augmentation. 
While white-box attacks provide theoretical understanding, it is the black-box attack—more emblematic of real-world situations—that presents substantial practical security challenges. Addressing this, our study introduces a transfer-based multimodal adversarial attack method. This method is capable of concurrently generating both adversarial texts and images with high transferability, thereby enabling a more comprehensive examination of the black-box adversarial perturbations within VLP systems 

% ---------------------------------------------------------------------------------------------------------------------------------

\section{Methodology }
\label{sec:method}
% In this section, we detail our proposed method: a gradient-based multimodal adversarial attack against vision-language per-training models. 
Our proposed method has two contributions:
(1) we optimize the perturbation and generate the multimodal adversarial samples simultaneously by unifying them into a gradient-based framework, which can better attack the vulnerable multimodal information that has similar semantics.
(2) we use contrastive learning to perturb the inherent structures in the benign samples and the contextual integrity of image-text pairs from various views, which can improve the transferability of the generated multimodal adversarial samples.

\begin{figure*}[ht]
\centering\includegraphics[width=6.0in, height=2.5in]{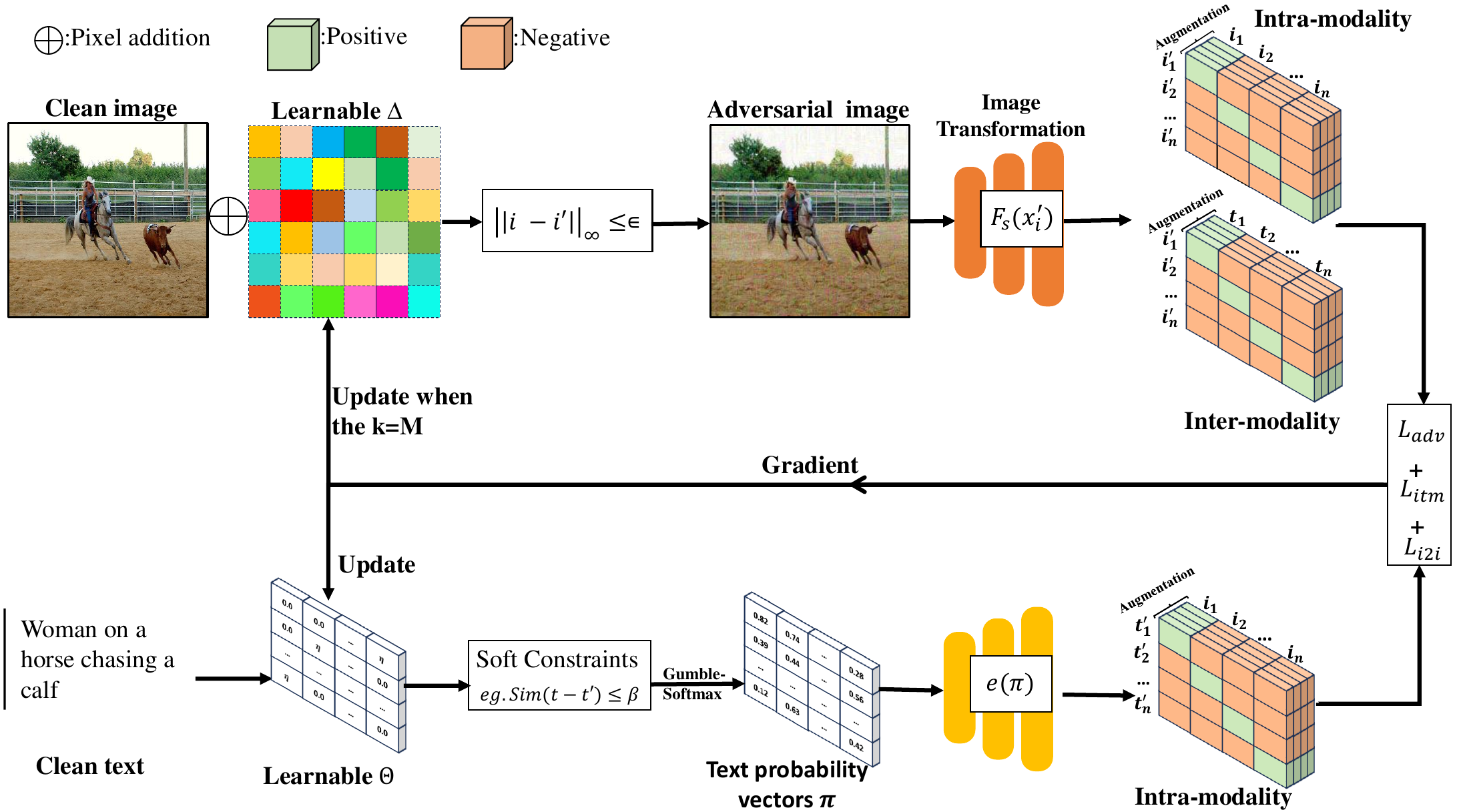}
\caption{Illustration of VLP-attack.  We propose a multimodal adversarial attack against VLP models, which is based on gradients to generate both adversarial text and adversarial images simultaneously, and leverages image-text contrastive loss and intra-modal contrastive loss to improve the transferability of multimodal adversarial samples. $\Delta$ is the perturbation for images and $\theta$ is the parameterized matrix for the Gumble-softmax sampling. $e(\cdot)$ is an embedding function. M is the number of image transformations. }
 \label{fig:framework}
\end{figure*}

\subsection{Problem Formulation}
\label{sec:L_adv}
Given an image-text pair, our goal is to minimize the semantic similarity between the image and the corresponding text with minor adversarial perturbations on images and texts on surrogate models.
Let $i$ denote an original image,  $t$ denote an original text, $i'$ denote an adversarial image, $t'$ denote an adversarial text, and $y$  denote the corresponding ground-truth label in the image-text retrieval task
and the visual entailment task.
We use $\mathcal{F}_s$  to denote the surrogate VLP models  and $\mathcal{F}_v$
to denote the victim VLP models, where the output of $\mathcal{F}_s$ is the feature representation of the input and the output of $\mathcal{F}_v$ is the prediction score.  
To generate valid multimodal adversarial samples, we need to minimize the cosine similarity Loss function $J$:

\begin{equation}
    L_{adv} = \left\{
\begin{array}{cl}
 \mathrm{min}\space J(\mathcal{F}_s(i'), \mathcal{F}_s(t'))\\
 s.t. ||i' - i||_{\infty} \leq \epsilon\\
 s.t. \mathrm{similarity}(t', t) \leq \beta\\
\end{array}
\right.
\end{equation}
% \yinpeng{the equation should be revised}
% \begin{equation}
%     L_{adv} = L(\mathcal{F_s}((X_i, X_t), Y^{truth}))
% \end{equation}
 on surrogate models
 % \yinpeng{what is J?}
 and make $\mathcal{F}_v((i', t')) = y^{adv}$ ($y \neq y^{adv}$) on victim models. For the image $i'$, under the constraint, the adding perturbation should look visually similar to the benign image $i$. For the text $t'$, we modify the words in a benign text $t$, which can not affect the original semantic understanding for humans. In this work, we use the $L_{\infty}$ norm to constrain the perceptibility of adversarial perturbations, i.e., $||i' - i||_{\infty} \leq \epsilon$. For the fluency of $t'$ , we constrain the $t'$ with the objective of the next token prediction by maximizing the likelihood given previous tokens. For the semantic similarity between $t$ and $t'$, we use the bert score to constrain the adversarial text $t'$. The framework of VLP-attack is shown in Figure~\ref{fig:framework}.

% % ------------------------------------------------------------------------------------------------------

\subsection{The objective for multimodal Adversarial Perturbation}
The optimization objective of multimodal adversarial perturbation consists of three components: image-text semantic similarity loss $L_{adv}$ for multimodal adversarial sample, soft constraints for adversarial texts, and contrastive loss for transferability, where $L_{adv}$ has been introduced in section~\ref{sec:L_adv}.

\subsubsection{Soft Constraints for Adversarial Texts}

\textbf{Adversarial Text Distribution.} Due to the discrete nature of text data, optimizing adversarial texts and images simultaneously requires transforming the text into a continuous distribution. Following the approach in Guo et al.\cite{guo2021gradient}, we employ the Gumbel-Softmax distribution\cite{jang2016categorical} to model text data. Specifically, for a sequence of words $t = [w_1, w_2, ..., w_n]$ with each $w_j$ belonging to a fixed vocabulary $\mathcal{V}$, we utilize a Gumbel-Softmax distribution $P_{\theta}$, parameterized by a matrix $\Theta \in \mathbb{R}^{n \times V}$. This distribution is used to sample a sequence of probability vectors $\pi$, where $\pi = \pi_1 ... \pi_n$ ($\pi _n$ denotes the probability vector of $w_n$ ). These vectors are then converted into input text embeddings using an embedding function $e(\cdot)$. Probability vectors $\pi$ and adversarial text features $e(\pi)$ are obtained according to the process:  
\begin{equation}
   (\pi _k )_j := \frac{\exp((\Theta _{k, j} + g_{k,j})/\tau)}{\sum _{v=1} ^V \exp((\Theta _{k,v} + g_{k,v})/\tau)}
\label{eq:gumble-softmax}
\end{equation}
\begin{equation}
    e(\pi) = e(\pi_1)\cdot \cdot \cdot e(\pi _n)
\end{equation}
 where  $g_{k,j}\sim $Gumble (0,1) and $\tau > 0$ is a temperature parameter that controls the smoothness of Gumble-softmax distribution. 

 \textbf{Soft Constraint.} For the fluency of the generated adversarial texts, we maximize the likelihood of the next token prediction when giving previous tokens:
 \begin{equation}
    L_{perp}(\pi) := - \sum^n_{k=1} \log\space p_{dis}(\pi _k|\pi_1 \cdot \cdot \cdot \pi _{k-1} )
\end{equation}
where $\mathrm{log}\space p_{dis}$  is the cross-entropy between the next word distribution and the previously predicted word distribution.
Meanwhile, to ensure that modifications in adversarial texts do not alter the original text's semantic interpretation for humans, we employ the BERT score~\cite{zhang2019bertscore} to maintain semantic consistency and constrain the semantic gap:
% for the semantic similarity between adversarial texts and origin texts, we use BERT score~\cite{zhang2019bertscore} to constraint the semantic gap:
\begin{equation}
    L_{sim}(t, t') = \sum_{k=1}^n w_k \underset{j=1,...,m}{\mathbf{max}} \phi(t)_k^\top \phi(t')_j
\end{equation}
where $t = t_1 \cdot \cdot \cdot t_n$ and  $t' = t'_1 \cdot \cdot \cdot t'_m$. $\phi(\cdot)$ is a language model that can generate contextualized embeddings for the token sequences.

\subsubsection{Contrastive Loss for Transferability}
 To improve the transferability of multimodal adversarial samples, we utilize contrastive learning including intra-modal contrastive learning and image-text contrastive learning to push the adversarial samples away from different views of the benign samples. Formally, the infoNCE loss for image-to-text is defined as:
\begin{equation}
    L_{nce}(i, t^+, t^- ) = E \left[ \log \frac{e^{(sim(i, t^+)/\tau)}}{\sum^K_{k=1} {e^{(sim(i, \hat t _k)/\tau)}}} \right]
\end{equation}
where $\tau$ is a temperature hyper-parameter, $\hat t$ is a set of all positive samples and negative samples for benign image $i$.

\textbf{Cross-modal contrastive Loss.} In image-text contrastive learning, our goal is to minimize the semantic similarity between the adversarial image and adversarial text that is matched in the original space. In image-to-text loss, the original text and text augmentations are taken as negative samples $t^-$.  In text-to-image loss, the original image and image augmentations are regarded as negative samples $i^-$. Intuitively, we encourage the features of the adversarial images and benign textual features to be unaligned in the embedding space, which can affect the multimodal features fusion in turn. Taken the image-to-text contrastive loss and text-to-image contrastive loss together as follows:
\begin{equation}
    L_{itm} = \frac{1}{2}\left[ L_{nce}(t', i^+, t^- ) + L_{nce}(i', t^+, i^-)\right]
\end{equation}
However, in addition to minimizing the semantic association between the adversarial images and the adversarial texts, we should also focus on the differences between the perturbed information and the original information in images.

\textbf{Intra-modal Contrastive Loss.} we utilize intra-modal contrastive learning to push adversarial samples semantically different from the benign samples in the same modality. Specifically, we consider some random views ($i_1, i_2, ..., i_n$) of the benign image $i$ under random data augmentation as negative samples $i^-$, and randomly sample images from the test set as the positive samples $i^+$.
% For the text modality, we apply round-trip translation~\cite{bayer2022survey} to construct the negative samples $x^-_t$ for the original text $x_t$. Overall, we minimize the following objective to encourage the adversarial samples away from the benign samples within one modality.
\begin{equation}
    L_{i2i} =  L_{nce}(i', i^+, i^-) 
\end{equation}

We utilize contrastive learning to discover and attack generalized patterns and structures in multimodal data so that multimodal adversarial samples can avoid trapping into model-specific local optimization making worse transferability of adversarial samples. 

\textbf{Objective Function. }We combine all the above losses into a final objective for the gradient-based multimodal adversarial samples generation :
\begin{equation}
    L = a \cdot L_{adv} + b \cdot L_{perp} + c \cdot L_{sim} + d \cdot L_{itm} +g \cdot L_{i2i}
    \label{eq:total_loss}
\end{equation}
where $a, b, c, d, g$ are the hyper-parameters that can control the strength of different constraint terms. Our purpose is to maximize the $L(\Theta, i')$ by optimizing the matrix $\Theta$ for adversarial text $t'$ and the perturbation on the image $i'$.

 \subsection{The generation of multimodal adversarial samples}
 We apply the gradient to optimize the adversarial texts and adversarial images simultaneously. For the adversarial texts, we optimize the matrix $\Theta$  as follows:
\begin{equation}
  \left\{
\begin{array}{cl}
 g^j_{\Theta} = \nabla _{\Theta}L(\mathcal{F}_s(i^j), e(\pi ^j)))\\
 \Theta ^{j+1} = f(\Theta ^j, \alpha_{\Theta}, g^j_{\Theta}, m^j_{\Theta})
\end{array}
\right.
\label{eq: update_texts_weights}
\end{equation}
 where $L$ is the total loss in eq (~\ref{eq:total_loss}),
 % \yinpeng{?},
 $\alpha_ \Theta$ is the learning rate for $\Theta$, $g^j_\Theta$ is the gradient of $\Theta$ at iteration $j$, $m^j_{\Theta}$ is the momentum at iteration $j$, and $f$ is the function of updating the weights of $\Theta$.
 
 For  the images that are continuous data, 
 similar to the momentum updating approach in~\cite{lin2019nesterov}, we use a variation of normal gradient descent to make a jump in the direction of previously accumulated gradients before computing the gradients in each iteration. Meanwhile,  following the~\cite{xie2019improving}, we diverse the pattern of $i^j$.
 % we can rotate\yinpeng{why?} $x_i$ by a small angle in  the direction of the gradient to produce imperceptible perturbation~\cite{goodfellow2014explaining}. 
 we input ($i', t'$) to $\mathcal{F}_s$ and obtain the gradient of $i'$ : 
 \begin{equation}
  \left\{
\begin{array}{cl}
 x_i^{nes} = T(i^j, p) + g_{i'}^j\\
 g^{j+1}_{i'} = \mu \cdot g^j_{i'} + \frac{\nabla _{i'} L(\mathcal{F}_s(i^{nes}), e(\pi ^j)))}{||\nabla _{i'} L(\mathcal{F}_s(i^{nes}), e(\pi^j))||_1}\\
 i^{j+1} = \mathrm{Clip}^{\epsilon} _{i}\{i^j + \alpha_{i'} \cdot \mathrm{sign}(g^{j+1}_{i'})\}\\
\end{array}
\right.
\label{eq:update_image}
\end{equation}
 where $T$ is the stochastic transformation function with the probability $p$, $\alpha _{i'}$ is the step size, $\mu$ is the decay factor of the momentum term, $g^j_{i'}$ is the accumulated gradient at iteration $j$, and $\mathrm{Clip}^{\epsilon}_{i}$ indicates the resulting adversarial images are clipped within the $\epsilon-$ball of the original image $i$.

Compared with the multimodal attack method based on the step-wise scheme~\cite{zhang2022towards, lu2023set}, we generate both the adversarial texts and the adversarial images simultaneously based on the gradient, which helps to find and attack the vulnerable multimodal information that has similar semantics. 
% The pseudocode of VLP-attack is outlined in Algorithm~\ref{alg: algorithm}.
The pseudocode of VLP-attack is outlined in Appendix~\uppercase\expandafter{\romannumeral6}.

\section{Experiments}
\label{sec:experiments}
In this section, we conduct a comprehensive evaluation of our proposed attack method across two key tasks: image-text retrieval and visual entailment. The objective is to demonstrate the efficacy of VLP-Attack on seven selected victim models. Additionally, we undertake ablation studies to substantiate the individual components of VLP-Attack and present a visualization of a pair of multimodal adversarial samples.
% We also provide a sensitivity analysis to shed light on the impact of various hyperparameters. 
% Concluding this section, we present a visualization of a pair of multimodal adversarial samples, followed by a discussion that contextualizes our findings within the broader scope of the work.  

% \usepackage{multirow}
\begin{table*}[ht]
\caption{Performance of transfer-based attack for image-text retrieval task on MSCOCO dataset and Flickr30K dataset.}
% \vspace{2mm}
\centering
\begin{tabular}{c|c|cccccc|cccccc}
\hline
\multirow{3}{*}{Model} & \multirow{3}{*}{Attack} & \multicolumn{6}{c|}{\begin{tabular}[c]{@{}c@{}}MSCOCO\\ (5K test set)\end{tabular}}           & \multicolumn{6}{c}{\begin{tabular}[c]{@{}c@{}}Flickr30K\\ (1K test set)\end{tabular}}         \\ \cline{3-14} 
                       &                         &               & TR            &               &               & IR            &               &               & TR            &               &               & IR            &               \\
                       &                         & R@1$\uparrow$           & R@5$\uparrow$           & R@10$\uparrow$          & R@1$\uparrow$           & R@5$\uparrow$           & R@10$\uparrow$          & R@1$\uparrow$           & R@5$\uparrow$           & R@10$\uparrow$          & R@1$\uparrow$           & R@5$\uparrow$           & R@10$\uparrow$          \\ \hline
\multirow{8}{*}{ALBEF} & BERT-Attack             & 15.3          & 8.9           & 5.6           & 19.5          & 18.1          & 15.0          & 7.7           & 0.7           & 0.4           & 17.1          & 9.9           & 7.2           \\
                       & GBDA                    & 14.7          & 13.0          & 11.1          & 11.1          & 15.7          & 16.2          & 6.5           & 1.6           & 1.2           & 14.7          & 13.2          & 12.5          \\
                       & TI-FGSM                 & 22.1          & 14.4          & 10.1          & 15.0          & 12.3          & 9.2           & 10.4          & 3.3           & 1.9           & 14.1          & 7.2           & 4.5           \\
                       & SI-NI-FGSM              & 26.7          & 19.8          & 14.7          & 20.0          & 18.2          & 14.7          & 22.9          & 12.0          & 8.9           & 24.1          & 14.9          & 10.8          \\
                       & AttackVLM               & 9.1           & 5.6           & 3.5           & 7.4           & 5.8           & 4.2           & 6.6           & 1.7           & 1.0           & 7.5           & 3.4           & 2.1           \\
                       & Co-attack               & 26.9          & 17.5          & 12.5          & 27.3          & 26.4          & 22.8          & 20.9          & 12.3          & 10.6          & 18.4          & 11.4          & 8.5           \\
                       & SGA                     & 34.3          & 26.0          & 19.4          & 30.6          & 29.8          & 25.7          & 27.9          & 12.1          & 9.2           & 32.4          & 21.4          & 16.4          \\
                       & \textbf{VLP-attack}     & \textbf{38.7} & \textbf{31.9} & \textbf{26.1} & \textbf{32.0} & \textbf{32.8} & \textbf{29.3} & \textbf{35.2} & \textbf{20.4} & \textbf{15.6} & \textbf{38.9} & \textbf{27.8} & \textbf{22.2} \\ \hline
\multirow{8}{*}{TCL}   & BERT-Attack             & 15.5          & 8.5           & 5.5           & 18.6          & 16.8          & 13.8          & 6.7           & 0.6           & 0.1           & 15.6          & 9.1           & 6.9           \\
                       & GBDA                    & 8.1           & 5.4           & 4.2           & 9.0           & 7.3           & 6.6           & 3.2           & 0.7           & 0.2           & 9.6           & 5.6           & 3.8           \\
                       & TI-FGSM                 & 14.2          & 8.5           & 5.6           & 9.4           & 7.6           & 5.4           & 22.9          & 11.7          & 7.9           & 24.7          & 15.0          & 10.8          \\
                       & SI-NI-FGSM              & 27.6          & 19.6          & 15.5          & 19.8          & 18.1          & 15.1          & 23.0          & 11.5          & 7.2           & 24.7          & 15.3          & 11.0          \\
                       & AttackVLM               & 12.3          & 7.8           & 5.3           & 9.9           & 8.4           & 6.5           & 12.3          & 7.8           & 5.3           & 9.8           & 8.4           & 6.5           \\
                       & Co-attack               & 27.4          & 19.2          & 14.2          & 27.7          & 27.2          & 23.8          & 15.5          & 5.4           & 1.7           & 26.4          & 17.5          & 13.4          \\
                       & SGA                     & 34.8          & 26.9          & 20.7          & 28.4          & 28.9          & 25.4          & 30.3          & 14.2          & 8.8           & 34.7          & 23.2          & 17.4          \\
                       & \textbf{VLP-attack}     & \textbf{36.2} & \textbf{29.0} & \textbf{23.7} & \textbf{29.1} & \textbf{29.6} & \textbf{26.1} & \textbf{32.3} & \textbf{17.2} & \textbf{12.2} & \textbf{37.4} & \textbf{26.1} & \textbf{20.2} \\ \hline
\multirow{8}{*}{BLIP}  & BERT-Attack             & 29.6          & 17.3          & 11.2          & 27.4          & 24.2          & 19.7          & 17.4          & 4.0           & 1.5           & 24.5          & 12.6          & 8.8           \\
                       & GBDA                    & 12.5          & 6.2           & 3.2           & 13.1          & 9.9           & 7.6           & 17.7          & 10.6          & 10.3          & 26.5          & 18.2          & 15.8          \\
                       & TI-FGSM                 & 24.8          & 13.5          & 8.6           & 17.5          & 12.6          & 9.5           & 30.2          & 13.5          & 9.1           & 27.7          & 15.5          & 10.8          \\
                       & SI-NI-FGSM              & 35.8          & 24.2          & 18.3          & 27.9          & 23.9          & 19.4          & 30.6          & 13.6          & 9.1           & 27.7          & 15.5          & 10.7          \\
                       & AttackVLM               & 11.1          & 6.8           & 4.3           & 10.3          & 8.5           & 6.7           & 4.7           & 0.8           & 0.5           & 6.3           & 2.7           & 1.8           \\
                       & Co-attack               & 40.1          & 27.6          & 19.3          & 35.5          & 33.2          & 28.3          & 26.0          & 7.7           & 4.1           & 31.4          & 18.7          & 13.2          \\
                       & SGA                     & 37.5          & 28.3          & 22.3          & 34.0          & 33.9          & 30.2          & 25.0          & 10.7          & 7.6           & 34.0          & 20.8          & 15.6          \\
                       & \textbf{VLP-attack}     & \textbf{49.0} & \textbf{38.4} & \textbf{29.8} & \textbf{39.2} & \textbf{39.0} & \textbf{34.3} & \textbf{44.7} & \textbf{23.3} & \textbf{15.3} & \textbf{43.6} & \textbf{30.4} & \textbf{23.3} \\ \hline
\multirow{8}{*}{BLIP2} & BERT-Attack             & 15.1          & 6.9           & 3.9           & 18.8          & 14.7          & 11.8          & 4.7           & 0.7           & 0.1           & 13.4          & 6.9           & 4.6           \\
                       & GBDA                    & 5.0           & 3.9           & 2.5           & 5.8           & 4.1           & 3.0           & 3.0           & 0.9           & 0.4           & 8.1           & 4.3           & 2.8           \\
                       & TI-FGSM                 & 5.6           & 2.5           & 1.3           & 4.2           & 3.2           & 1.9           & 3.1           & 0.2           & 1.8           & 3.6           & 1.2           & 0.7           \\
                       & SI-NI-FGSM              & 12.7          & 6.5           & 4.2           & 9.4           & 7.4           & 6.0           & 6.1           & 1.2           & 1.0           & 7.4           & 3.3           & 2.2           \\
                       & AttackVLM               & 4.4           & 2.0           & 1.1           & 4.0           & 2.7           & 2.1           & 2.8           & 0.4           & 0.2           & 3.3           & 1.1           & 0.8           \\
                       & Co-attack               & 19.6          & 9.7           & 6.2           & 20.3          & 17.8          & 14.2          & 6.5           & 1.5           & 0.5           & 17.3          & 9.6           & 6.7           \\
                       & SGA                     & 19.9          & 10.6          & 7.2           & 20.2          & 17.7          & 14.6          & 10.7          & 2.0           & 0.8           & 18.2          & 9.9           & 7.1           \\
                       & \textbf{VLP-attack}     & \textbf{22.0} & \textbf{12.7} & \textbf{8.6}  & \textbf{20.9} & \textbf{17.9} & \textbf{14.9} & \textbf{11.5} & \textbf{3.2}  & \textbf{2.1}  & \textbf{20.5} & \textbf{11.3} & \textbf{8.3}  \\ \hline
\end{tabular}
\label{tab:  image-text-retrieval}
\begin{threeparttable}
 TR means Text Retrieval, IR means Image Retrieval 
\end{threeparttable}
\end{table*}

\begin{table*}[!ht]
\caption{Performance of transfer-based attack for visual entailment task on SNLI-VE dataset.}
\centering
\begin{tabular}{c|cccccccc}
\hline
          \diagbox{Attack$\uparrow$}{Model}& BERT-Attack & GBDA & TI-FGSM & SI-NI-FGSM & AttackVLM & Co-attack & SGA  & \textbf{VLP-attack} \\ \hline
ALBEF     & 17.6        & 20.2 & 10.5    & 11.8       & 2.6       & 22.1      & 25.5 & \textbf{28.9}       \\ \hline
TCL       & 12.1        & 15.2 & 8.7     & 10.2       & 1.4       & 21.8      & 23.4 & \textbf{27.0}       \\ \hline
BLIP2     & 7.2         & 16.2 & 9.1     & 16.9       & 8.6       & 9.8       & 18.4 & \textbf{22.0}       \\ \hline
MiniGPT-4 & 2.3         & 11.4 & 6.1     & 4.4        & 4.6       & 6.1       & 12.0 & \textbf{12.9}       \\ \hline
\end{tabular}
\label{tab:SNLI-VE}
\end{table*}

% \usepackage{multirow}
\begin{table*}[ht]
\caption{Evaluation of the proposed methods on CLIP series models on image-text retrieval tasks.}
% \vspace{2mm}
\centering
\begin{tabular}{c|c|cccccc|cccccc}
\hline
\multirow{3}{*}{Model}             & \multirow{3}{*}{Attack} & \multicolumn{6}{c|}{\begin{tabular}[c]{@{}c@{}}MSCOCO\\ (5K test set)\end{tabular}}           & \multicolumn{6}{c}{\begin{tabular}[c]{@{}c@{}}Flickr30K\\ (1K test set)\end{tabular}}         \\ \cline{3-14} 
                                   &                         &               & TR            &               &               & IR            &               &               & TR            &               &               & IR            &               \\
                                   &                         & R@1$\uparrow$           & R@5$\uparrow$           & R@10$\uparrow$          & R@1$\uparrow$           & R@5$\uparrow$           & R@10$\uparrow$          & R@1$\uparrow$           & R@5$\uparrow$           & R@10$\uparrow$          & R@1$\uparrow$           & R@5$\uparrow$           & R@10$\uparrow$          \\ \hline
\multirow{4}{*}{\makecell[c]{Open CLIP\\-G/14}}    & AttackVLM               & 9.1           & 5.6           & 3.5           & 7.4           & 5.8           & 4.2           & 6.6           & 1.7           & 1.0           & 7.5           & 3.4           & 2.1           \\
                                   & Co-attack               & 26.6          & 22.3          & 18.2          & 18.8          & 19.0          & 17.0          & 22.8          & 6.6           & 2.8           & 26.1          & 18.7          & 13.7          \\
                                   & SGA                     & 27.4          & 22.4          & 18.4          & 19.2          & 18.8          & 17.4          & 24.2          & 9.9           & 5.3           & 27.8          & 19.1          & 14.0          \\
                                   & \textbf{VLP-attack}     & \textbf{28.1} & \textbf{22.7} & \textbf{18.9} & \textbf{19.5} & \textbf{20.8} & \textbf{18.4} & \textbf{32.2} & \textbf{16.7} & \textbf{11.4} & \textbf{33.3} & \textbf{24.1} & \textbf{18.1} \\ \hline
\multirow{4}{*}{\makecell[c]{EVA-02-CLIP\\bigE-14-plus}} & AttackVLM               & 17.9          & 10.6          & 7.4           & 2.4           & 2.0           & 1.7           & 10.3          & 2.5           & 2.3           & 3.7           & 3.1           & 2.8           \\
                                   & Co-attack               & 20.2          & 17.4          & 13.6          & 18.0          & 18.2          & 16.6          & 20.6          & 4.1           & 1.7           & \textbf{24.2}          & \textbf{17.1 }         & \textbf{13.1}          \\
                                   & SGA                     & 23.6          & 19.7          & 15.8          & 19.3          & 19.9          & 18.1          & 23.1          & 5.0           & 2.6           & 21.7          & 15.5          & 11.3          \\
                                   & \textbf{VLP-attack}     & \textbf{29.7} & \textbf{20.6} & \textbf{15.2} & \textbf{21.0} & \textbf{21.6} & \textbf{19.4} & \textbf{27.9} & \textbf{9.5}  & \textbf{5.6}  & 23.5 & 16.0 & 12.1 \\ \hline
\end{tabular}
\label{tab:CLIP-series-image-text-retrieval}
\begin{threeparttable}
 TR means Text Retrieval, IR means Image Retrieval 
\end{threeparttable}
\end{table*}

% \usepackage{multirow}
\begin{table*}[ht]
\caption{The Semantic similarity (Sim.) and Token error rate (TER.) of adversarial texts for all the attacks.}
\centering
\begin{tabular}{c|c|c|ccccc}
\hline
Tasks                                 & Datasets                   & Metrics & Bert-Attack & GBDA & Co-attack & SGA  & VLP-attack \\ \hline
\multirow{4}{*}{Image-text Retrieval} & \multirow{2}{*}{MSCOCO}    & Sim.$\uparrow$    & 89.9        & 82.5 & 89.8      & 90.0 & 90.1       \\
                                      &                            & TER.$\downarrow$    & 12.3        & 35.6 & 12.3      & 12.3 & 14.3       \\ \cline{2-8} 
                                      & \multirow{2}{*}{Flickr30K} & Sim.$\uparrow$    & 90.7        & 84.6 & 90.7      & 90.4 & 90.3       \\
                                      &                            & TER.$\downarrow$    & 10.8        & 33.4 & 10.8      & 10.8 & 14.6       \\ \hline
\multirow{2}{*}{Visual Entailment}    & \multirow{2}{*}{SNLI-VE}   & Sim.$\uparrow$    & 85.7        & 85.6 & 85.7      & 85.8 & 85.1       \\
                                      &                            & TER.$\downarrow$    & 18.1        & 24.1 & 18.1      & 18.1 & 20.9       \\ \hline
\end{tabular}
\label{tab: text_similarity}
\end{table*}

\label{sec:experiments}
\subsection{Experimental Settings}
\textbf{Datasets and Downstream Tasks.}
Following Co-attack~\cite{zhang2022towards}, we use the test set in MSCOCO~\cite{lin2014microsoft}, Flickr30K~\cite{plummer2015flickr30k}, SNLI-VE~\cite{xie2019visual} to evaluate the effectiveness of our proposed method at different tasks in a black-box setting, where MSCOCO and Flickr30K are used to evaluate the Text Retrieval (TR) and Image Retrieval (IR) tasks, and SNLI-VE is used to evaluate the Visual Entailment (VE) task. Note that due to the nature of the adversarial attack, we select only the image-text pairs with a label of entailment from the test set of the SNLI-VE dataset for the VE task.

\textbf{Evaluation Metrics.}
For evaluation, we report (1) the attack success rate (ASR), which measures the rate of successful adversarial samples.\textbf{ Similar to Co-attack~\cite{zhang2022towards}, we define the ASR as the results when the original samples are fed into the target models minus the results when the adversarial samples are fed into the target models.}
(2) Semantic Similarity (Sim.), which is computed between the original and
adversarial sentences and is commonly approximated by the universal sentence encoder~\cite{yang2019multilingual}. 
(3) Tokens error rate (TER.), which is the proportion of words or characters in a text that have been altered to create an adversarial text, generally, less perturbation results in more semantic consistency. 
Note: Semantic Similarity (Sim.) and Token Error Rate (TER) are used to evaluate the extent of change between original and adversarial texts, identifying methods with superior ASR in a black-box setting under equivalent semantic deviation and token errors. 
The whole method is implemented by Pytorch~\cite{paszke2019pytorch} and all experiments are conducted in a GeForce RTX A6000 GPU.

\textbf{Hyperparameters.}
For a fair comparison, 
the maximum perturbation $\epsilon$ is set to 16/255 among all experiments for the adversarial attack on images.  The step size is set to $\epsilon / 10$. The number of iterations for all image attacks is 10 except for AttackVLM which is 100 according to the setting in the original paper.  For the adversarial attack on texts, the max length of each text is set to 30.
The adversarial distribution parameter $\Theta$ is optimized using RMSProp~\cite{zeiler2012adadelta} with a learning rate of 0.3 and a momentum of 0.6. For the distribution parameters, $\Theta$ is initialized to zero except $\Theta _{j,k} = \eta$ where $j$ denotes the $j_{th}$ word of the benign text, and $k$ denotes the id of the word. In the experiment, we take $\eta = 13$.
For the text augmentation, we use the round-trip translation skill to produce diverse texts.
The transformation operations $T$ consist of rotation, polarizing, translation, shear, color jittering, and cropping.
The number of images transformation is set to 5, and the probability $p$ is 0.6. 
The number of data augmentation is set to 7. 
For the image-text retrieval task, $a=8,  b=c=d=g=1$. For the visual entailment task, $a=10, b=c=d=g=1$.

\textbf{Vision-Language Pre-training Models.}
Vision-language representation learning largely benefits from image-text alignment through contrastive losses. The VLP models examined in our work are as follows:
\begin{itemize}
\item \textbf{CLIP}~\cite{radford2021learning}
% is a pre-training VLP model on 400 million (image, text) pairs collected from the internet, 
can understand images and texts simultaneously through image-text contrastive learning, which has been a general paradigm to align the image and text representations in latest VLP models. In this work, we utilize CLIP as the surrogate model to craft the black-box attacks to explore the transferability of multimodal adversarial examples for VLP models and take the \textbf{ViT-B/16} as the image encoder.
\item \textbf{Open CLIP-G/14}~\cite{ilharco_gabriel_2021_5143773} utilizes the LAION-2B English subset from the LAION-5B collection for training, employing the Open CLIP framework. Compared to OpenAI's CLIP-B/16, OpenCLIP-G/14 features an expanded parameter set and a larger training dataset. 
\item \textbf{EVA-02-CLIP-bigE-14-plus}~\cite{sun2023eva} integrates advancements in representation learning, optimization, and augmentation, significantly outperforming OpenAI's CLIP models with equivalent parameter sizes. In comparison to OpenAI's CLIP-B/16, EVA-02-CLIP-bigE-14-plus benefits from an expanded dataset and increased parameter count, enhancing its performance. 
\item \textbf{ALBEF}~\cite{li2021align} aligns the image and text representation by the image-text contrastive loss before Fusing through cross-attention, which demonstrates the effectiveness on various V+L tasks including image-text retrieval, and visual entailment. 
 % In the experiment, we use an ALBEF version that has been fine-tuned on the respective datasets. 
\item \textbf{TCL}~\cite{yang2022vision} proposes triple contrastive learning for VLP by leveraging both cross-modal and intra-modal self-supervision and firstly takes into account local structure information for multi-modality representation. 
% In the experiment, we use a TCL version that has been fine-tuned on the respective datasets.
\item \textbf{BLIP}~\cite{li2022blip} utilizes a dataset bootstrapped from large-scale noisy image-text pairs by injecting diverse synthetic captions and removing noisy captions to pre-train a multimodal encoder-decoder model. The image-text contrastive loss is utilized to align the feature space of the visual transformer and text transformer. Due to the absence of fine-tuned models for the SNLI-VE dataset provided by BLIP, we restricted our utilization of the BLIP model solely to the image-text retrieval task. 
\item \textbf{BLIP2}~\cite{li2023blip} 
% employs a lightweight Querying Transformer to navigate the disparity between modalities. 
facilitates the extraction of visual representations that resonate most closely with the textual context. Concurrently, through image-text contrastive learning, there is an alignment of text and image representations. For the image-text retrieval task, we use the BLIP2~\footnote{\href{https://github.com/salesforce/LAVIS/tree/main/projects/blip2}{LAVIS/projects/blip2 at main · salesforce/LAVIS (github.com)} } finetuned on the MSCOCO dataset. For the visual entailment task, we use the blip2-opt-2.7b~\footnote{\href{https://huggingface.co/Salesforce/blip2-opt-2.7b}{Salesforce/blip2-opt-2.7b · Hugging Face} }.
\item \textbf{MiniGPT-4}~\cite{zhu2023minigpt} aligns the visual features with a large language model and can be used for image-grounded text generation tasks. In the experiments,  we use Vicuna-7B~\footnote{\href{https://huggingface.co/Vision-CAIR/vicuna-7b/tree/main}{Vision-CAIR/vicuna-7b at main (huggingface.co)} } for the visual entailment task. 
\end{itemize}
 
\textbf{Transfer-based Adversarial Attack Baselines.}
To demonstrate the effectiveness of our proposed method, in the transfer-based attack setting, we select seven attacks as the baseline methods.
\begin{itemize}
    \item \textbf{Bert-attack}~\cite{li2020bert} generates adversarial texts that can deceive NLP models while maintaining semantic similarity and grammatical correctness through pre-trained masked language models exemplified by BERT. 
    \item \textbf{GBDA}~\cite{guo2021gradient} incorporates differentiable fluency and semantic similarity constraints to the adversarial loss and generates the adversarial texts by optimizing a parameterized distribution. For a fair comparison, instead of using GBDA to generate adversarial text by querying against the unknown target model, we sampled the adversarial texts directly from the optimized adversarial distribution.
    \item \textbf{TI-FGSM}~\cite{dong2019evading} is an extension of FGSM~\cite{goodfellow2014explaining}, which improves upon the FGSM by introducing translation invariance to increase the transferability of adversarial examples between different models. 
    \item \textbf{SI-NI-FGSM}~\cite{lin2019nesterov} adapts the Nesterov accelerated gradient method to jump out of the local optimal solution and utilizes the scale-invariant property to improve the transferability of adversarial images.
    \item \textbf{AttackVLM}~\cite{zhao2023evaluating} is a transfer-based adversarial attack that perturbs the images to evaluate the adversarial robustness of large vision-language models. According to the performance of black-box attack against victim models and settings, we select Matching image-image features (MF-ii) in it. 
    \item \textbf{Co-attack}~\cite{zhang2022towards} is a multimodal adversarial attack against vision-language pre-training models that adopts a step-wise mechanism that first perturbs the discrete inputs (text) and then perturbs the continuous inputs (image) given the text perturbation, which is designed for white-box attack manner.
    \item \textbf{SGA}~\cite{lu2023set} is a multimodal transfer-based adversarial attack that investigates the adversarial transferability of recent VLP models through modalities interactions and data augmentation.
\end{itemize}

%\subsubsection{Transferability Against VLP Models}
\subsection{Main Results}
The performance of the adversarial samples generated by our proposed method and other baseline methods against  VLP models  for the image-text retrieval task and the visual entailment task is reported in Table~\ref{tab: image-text-retrieval}, Table~\ref{tab:SNLI-VE}, Table~\ref{tab:CLIP-series-image-text-retrieval} and Table~\ref{tab: text_similarity}. We have several observations:
\begin{itemize}
    \item Our proposed VLP-attack, consistently exhibits competitive or superior performance compared to other baseline methods in text-image retrieval and visual entailment tasks, except for a slightly lower performance than Co-attack against EVA-02-CLIP-BigE-14-plus in the Flickr30K, as shown in Table~\ref{tab:CLIP-series-image-text-retrieval}. This demonstrates the effectiveness of the VLP-attack.
    % \item Compared with Co-attack, our proposed method has a significantly higher attack success rate in a transfer-based setting at a close similarity and the tokens error rate of adversarial texts, even though both methods attack images and texts as the inputs to influence the output of the target models. The reason is that contrastive learning in  VLP-attack can destroy more low-level and mid-level information on the image-text pairs and mitigate the overfitting of target models to improve the transferability of adversarial samples than Co-attack with a step-wise scheme to attack multimodal data.
    \item  Compared with single-modal transfer attack methods in Table~\ref{tab: image-text-retrieval} and Table~\ref{tab:SNLI-VE}, the multimodal adversarial attack method leverages the joint information from multiple modalities (such as images and texts), the attack can exploit the vulnerabilities and inconsistencies across different modalities, making the attack more effective. 
    Nonetheless, due to Co-attack being designed for a white-box setting, the attack success rate of multimodal adversarial samples is lower than transfer-based image adversarial attack methods when attacking TCL and BLIP in Flickr30K dataset.
    \item For the BLIP2 model, the performance of VLP-attack seems a bit more modest, particularly for Text Retrieval. However, in Table~\ref{tab: different_iteration}~\footnote{To avoid the effect of step size, we set the step size $\alpha$ to 0.1 when the number of iterations is between 20-50, and modify the learning rate to ensure the Sim. and TER. rate closed with that when the number of iterations is 10.}, we observe that as the number of iterations increases, VLP-attack generally demonstrates a positive trend in performance, which indicates that VLP-attack may benefit from more iterations, optimizing the adversarial perturbations.
    \item Compared to SGA, Co-attack and Bert-attack, the VLP-attack has a slightly higher TER and Similarity as shown in Table~\ref{tab: text_similarity}. This suggests that while VLP-attack might introduce more token-level perturbations, it doesn't compromise the overall semantic integrity of the adversarial texts.
    \item The VLP-attack, as evidenced by the experimental results, stands out as a potent method in the realm of transfer-based multimodal attacks for perturbing the understanding between images and texts in VLP models. It not only showcases impressive performance metrics across different models but also ensures that the generated adversarial samples preserve their semantic content.
\end{itemize}

\subsection{Ablation Study}

% \usepackage{multirow}
\begin{table}[ht]
\caption{Evaluation of the proposed methods on SNLI-VE dataset for visual entailment task.}
% \vspace{2mm}
\centering
\begin{tabular}{ccccc}
\hline
\multicolumn{5}{c}{SNLI-VE}                                                                                                                              \\ \hline
\multicolumn{1}{c|}{Model}                                     & \multicolumn{1}{c|}{Attack}             & ASR$\uparrow$         & Sim.$\uparrow$          & TER.$\downarrow$          \\ \hline
\multicolumn{1}{l|}{\multirow{4}{*}{ALBEF}} & \multicolumn{1}{c|}{VLP-attack@i2i+itm} & 24.5          & 85.7          & 20.5          \\
\multicolumn{1}{c|}{}                                          & \multicolumn{1}{c|}{VLP-attack@i2i}     & 28.1          & \textbf{85.8}          & \textbf{20.2}          \\
\multicolumn{1}{c|}{}                                          & \multicolumn{1}{c|}{VLP-attack@itm}     & 25.9          & 85.6          & 20.7          \\
\multicolumn{1}{c|}{}                                          & \multicolumn{1}{c|}{VLP-attack}         & \textbf{28.9} & 85.1 & 20.9 \\ \hline
\multicolumn{1}{l|}{\multirow{4}{*}{TCL}}   & \multicolumn{1}{c|}{VLP-attack@i2i+itm} & 20.1          & 85.7          & 20.5          \\
\multicolumn{1}{l|}{}                                          & \multicolumn{1}{c|}{VLP-attack@i2i}     & 26.2          & \textbf{85.8}          & \textbf{20.2}          \\
\multicolumn{1}{l|}{}                                          & \multicolumn{1}{c|}{VLP-attack@itm}     & 25.0          & 85.6          & 20.7          \\
\multicolumn{1}{l|}{}                                          & \multicolumn{1}{c|}{VLP-attack}         & \textbf{27.0} & 85.1 & 20.9 \\ \hline
\multicolumn{1}{l|}{\multirow{4}{*}{BLIP2}}   & \multicolumn{1}{c|}{VLP-attack@i2i+itm} & 18.6          & 85.7          & 20.5          \\
\multicolumn{1}{l|}{}                                          & \multicolumn{1}{c|}{VLP-attack@i2i}     & 19.7          & \textbf{85.8}          & \textbf{20.2}          \\
\multicolumn{1}{l|}{}                                          & \multicolumn{1}{c|}{VLP-attack@itm}     & 21.1          & 85.6          & 20.7          \\
\multicolumn{1}{l|}{}                                          & \multicolumn{1}{c|}{VLP-attack}         & \textbf{22.0} & 85.1 & 20.9 \\ \hline
\multicolumn{1}{l|}{\multirow{4}{*}{MiniGPT-4}}   & \multicolumn{1}{c|}{VLP-attack@i2i+itm} & 10.8          & 85.7          & 20.5          \\
\multicolumn{1}{l|}{}                                          & \multicolumn{1}{c|}{VLP-attack@i2i}     & 11.3          & \textbf{85.8}          & \textbf{20.2}          \\
\multicolumn{1}{l|}{}                                          & \multicolumn{1}{c|}{VLP-attack@itm}     & 11.4          & 85.6          & 20.7          \\
\multicolumn{1}{l|}{}                                          & \multicolumn{1}{c|}{VLP-attack}         & \textbf{12.9} & 85.1 & 20.9 \\ \hline
\end{tabular}
 
\label{tab:VE_ablation}
\end{table}

% \usepackage{multirow}
\begin{table}[ht]
\caption{Evaluation of the proposed methods on Flickr30k dataset for image-text retrieval task. }
% \vspace{2mm}
\centering
\begin{threeparttable}
\begin{tabular}{cccccc}
\hline
\multicolumn{6}{c}{Flickr30k (1K test set)}                                                                                                                                           \\ \hline
\multicolumn{1}{c|}{Model}                                     & \multicolumn{1}{c|}{Attack}             & TR*$\uparrow$            & IR*$\uparrow$            & Sim.$\uparrow$          & TER.$\downarrow$          \\ \hline
\multicolumn{1}{c|}{\multirow{4}{*}{ALBEF}} & \multicolumn{1}{c|}{VLP-attack@i2i+itm} & 27.6          & 20.3          & 90.4          & 14.4          \\
\multicolumn{1}{c|}{}                                          & \multicolumn{1}{c|}{VLP-attack@i2i}     & 31.0          & 32.6          & \textbf{90.5}          & \textbf{14.4}          \\
\multicolumn{1}{c|}{}                                          & \multicolumn{1}{c|}{VLP-attack@itm}     & 30.6          & 33.1          & \textbf{90.5}          & \textbf{14.4}          \\
\multicolumn{1}{c|}{}                                          & \multicolumn{1}{c|}{VLP-attack}         & \textbf{35.2} & \textbf{38.9} & 90.3 & 14.6 \\ \hline
\multicolumn{1}{c|}{\multirow{4}{*}{TCL}}   & \multicolumn{1}{c|}{VLP-attack@i2i+itm} & 14.5          & 19.6          & 90.4          & 14.4          \\
\multicolumn{1}{c|}{}                                          & \multicolumn{1}{c|}{VLP-attack@i2i}     & 30.1          & 34.7          & \textbf{90.5}          & \textbf{14.4}          \\
\multicolumn{1}{c|}{}                                          & \multicolumn{1}{c|}{VLP-attack@itm}     & 29.4          & 31.6          & \textbf{90.5}          & \textbf{14.4}          \\
\multicolumn{1}{c|}{}                                          & \multicolumn{1}{c|}{VLP-attack}         & \textbf{32.3} & \textbf{37.4} & 90.3 & 14.6 \\ \hline
\multicolumn{1}{c|}{\multirow{4}{*}{BLIP}}  & \multicolumn{1}{c|}{VLP-attack@i2i+itm} & 28.6          & 37.2          & 90.4          & 14.4          \\
\multicolumn{1}{c|}{}                                          & \multicolumn{1}{c|}{VLP-attack@i2i}     & 41.8          & 41.4          & \textbf{90.5}          & \textbf{14.4}          \\
\multicolumn{1}{c|}{}                                          & \multicolumn{1}{c|}{VLP-attack@itm}     & 37.4          & 38.2          & \textbf{90.5}          & \textbf{14.4}          \\
\multicolumn{1}{c|}{}                                          & \multicolumn{1}{c|}{VLP-attack}         & \textbf{44.7} & \textbf{43.6} & 90.3 & 14.6 \\ \hline
 \multicolumn{1}{c|}{\multirow{4}{*}{BLIP2}}  & \multicolumn{1}{c|}{VLP-attack@i2i+itm} & 5.5& 11.5& 90.4          & 14.4          \\
\multicolumn{1}{c|}{}                                          & \multicolumn{1}{c|}{VLP-attack@i2i}     & 6.9& 14.0& \textbf{90.5}& \textbf{14.4}\\
\multicolumn{1}{c|}{}                                          & \multicolumn{1}{c|}{VLP-attack@itm}     & 10.1& 18.9& \textbf{90.5}& \textbf{14.4}\\
\multicolumn{1}{c|}{}                                          & \multicolumn{1}{c|}{VLP-attack}         & \textbf{11.5}& \textbf{20.5}& 90.3 & 14.6 \\ \hline
\end{tabular}
 \begin{tablenotes}
        \footnotesize
        \item * means the attack success rate of R@1 is reported.
      \end{tablenotes}
  \end{threeparttable}

 \label{tab:retrieval_ablation}
\end{table}

\begin{table}[ht]
\caption{The fluency and similarity evaluation of the proposed methods against ALBEF for image-text retrieval task.}
\vspace{2mm}
\centering
\begin{threeparttable}
\begin{tabular}{cccccc}
\hline
\multicolumn{6}{c}{Flickr30k (1K test set)}                                                                             \\ \hline
\multicolumn{1}{c|}{Attack}          & TR*$\uparrow$            & IR*$\uparrow$            & Sim.$\uparrow$          & TER.$\downarrow$          & Perp. $\downarrow$    \\ \hline
\multicolumn{1}{c|}{VLP-attack@sim}  & \textbf{36.0}          & 38.9          & 88.6          & 14.3          & 62.3          \\
\multicolumn{1}{c|}{VLP-attack@perp} & 33.6          & 37.5          & 90.0          & \textbf{14.0}          & 62.5          \\
\multicolumn{1}{c|}{VLP-attack}      & 35.2 & \textbf{38.9} & \textbf{90.3} & 14.6 & \textbf{61.6} \\ \hline
\end{tabular}
\begin{tablenotes}
        \footnotesize
        \item  * means the attack success rate of R@1 is reported.
      \end{tablenotes}
  \end{threeparttable}
  \label{tab:ablation_smi_fluency}
\end{table}

% \usepackage{multirow}
\begin{table*}[ht]
\caption{Performance of VLP-attack attack with the different number of iterations against BLIP2 for image-text retrieval task}
\vspace{2mm}
\centering
\begin{tabular}{ccccccccc}

\hline
\multicolumn{9}{c}{Flickr30K (1K test set)}                                                                                                                                                                                        \\ \hline
\multicolumn{1}{c|}{\multirow{2}{*}{\begin{tabular}[c]{@{}c@{}}Number of\\ Iteration\end{tabular}}} & \multicolumn{3}{c}{Text Retrieval} & \multicolumn{3}{c}{Image Retrieval} & \multirow{2}{*}{Sim.$\uparrow$} & \multirow{2}{*}{TER.$\downarrow$} \\ \cline{2-7}
\multicolumn{1}{c|}{}                                                                               & R@1$\uparrow$        & R@5$\uparrow$       & R@10$\uparrow$      & R@1$\uparrow$        & R@5$\uparrow$        & R@10$\uparrow$      &                       &                       \\ \hline
\multicolumn{1}{c|}{10}                                                                             & 11.5       & 3.2       & 2.0       & 20.5       & 11.3       & 8.2       & 90.3                  & 14.6                  \\ \hline
\multicolumn{1}{c|}{20}                                                                             & 19.7       & 7.2       & 5.7       & 25.2       & 15.3       & 11.1      & 91.9                  & 12.3                  \\ \hline
\multicolumn{1}{c|}{30}                                                                             & 21.5       & 8.4       & 5.3       & 28.9       & 18.1       & 13.4      & 90.3                  & 14.1                  \\ \hline
\multicolumn{1}{c|}{40}                                                                             & 22.1       & 10.1      & 6.5       & 30.3       & 18.5       & 14.1      & 90.3                  & 14.6                  \\ \hline
\multicolumn{1}{c|}{50}                                                                             & 25.4       & 12.7      & 7.4       & 31.9       & 20.9       & 15.8      & 89.7                  & 15.5                  \\ \hline
\end{tabular}
% \begin{threeparttable}
% To avoid the effect of step size, we set the step size $\alpha$ to 0.1 when the number of iterations is between 20-50,\\ and modify the learning rate to ensure the Sim. and TER. rate closed with that when the number of iterations is 10.
% \end{threeparttable}
\label{tab: different_iteration}
\end{table*}

\subsubsection{Ablation study on contrastive learning}
We conducted ablation experiments to study the impact of contrastive learning on image-text retrieval tasks and visual entailment tasks. To demonstrate the effectiveness of VLP-attack, we compare variants of VLP-attack with respect to the following perspectives: (1) the effect of image-text contrastive learning, (2) the effect of intra-modal contrastive learning. The following VLP-attack variants are designed for comparison.
\begin{itemize}
    \item VLP-attack@i2i: A variant of VLP-attack with the intra-modal contrastive learning being removed.
    \item VLP-attack@itm: A variant of VLP-attack with the image-text contrastive learning being removed.
    \item VLP-attack@i2i+itm: A variant of VLP-attack with the intra-modal contrastive learning and image-text contrastive learning being removed.
\end{itemize}
The ablation study results are shown in Table~\ref{tab:VE_ablation} and Table~\ref{tab:retrieval_ablation}. For the image-text retrieval task and visual entailment task, we can have the following observations: 
\begin{itemize}
    \item The attack success rate (ASR) has a significant drop when removing the contrastive learning, but the similarity (Sim.) and tokens error rate (TER.) change a little, which suggests that contrastive learning  contributes to the transferability of multimodal adversarial samples. 
    \item Contrastive learning including image-text contrastive learning and intra-modal contrastive learning all have a positive impact on the transferability of adversarial samples, where the reason is that self-supervised contrastive learning forces the perturbation to focus on the inherent structures in the benign samples from different views and ignore irrelevant factors or nuisances. 
    \item A greater drop in ASR for the image-text retrieval task compared to the visual entailment task when removing the contrastive learning from VLP-attack may be due to the perturbation crafted by contrastive learning can have a  greater effect on the alignment for image-text pairs, which is more important in image-text retrieval task.
\end{itemize}

\subsubsection{Ablation study on fluency and similarity of adversarial texts}
We conducted ablation experiments to study the impact of similarity constraints and fluency constraints for adversarial texts in Flickr30k dataset for the image-text retrieval task. We use the Perplexity score (Perp.) to evaluate the fluency of adversarial texts, which is computed with the perplexity score of GPT-2 (large). 
The following VLP-attack variants are designed for comparison:
\begin{itemize}
    \item VLP-attack@perp: A variant of VLP-attack with the fluency constraint being removed.
    \item VLP-attack@sim: A variant of VLP-attack with the similarity constraint being removed.
\end{itemize}
The ablation study results are shown in Table~\ref{tab:ablation_smi_fluency}. We can observe that the 
similarity (Sim.) is reduced from 90.3 to 88.6 and Perplexity (Perp.) is increased from 61.6 to 62.3  when removing the similarity constraints from VLP-attack, which suggests that the similarity constraints contribute to the preservation of adversarial texts' semantics. For the perplexity constraints, the R@1 in text retrieval (TR), R@1 in image retrieval (IR), and Perp. show obvious change when removing perplexity constraints from VLP-attack, indicating the perplexity constraints can improve the attack success rate and fluency of adversarial texts generated by our method.

Ablation experiments were carried out to assess the influence of contrastive learning, similarity constraints, and fluency constraints in VLP-attack. Our findings reveal that each component in VLP-attack significantly contributes to generating effective multimodal adversarial samples. 
% Complete details and analysis are provided in Appendix A.2. 

% ---------------------------------------------------------------------------------------------------------------
\subsection{Sensitivity Analysis }
To clarify the selection of hyperparameters $a, b, c, d, g$, as well as the choices for data augmentation frequency and number of image transformations in the VLP-attack, we conducted a comprehensive sensitivity analysis.

\subsubsection{Hyperparameter Selection for $a, b, c, d, g$} 
\begin{figure*}[ht]
\centering
\captionsetup[subfloat]{}
\subfloat[]{\includegraphics[width=0.45\textwidth]{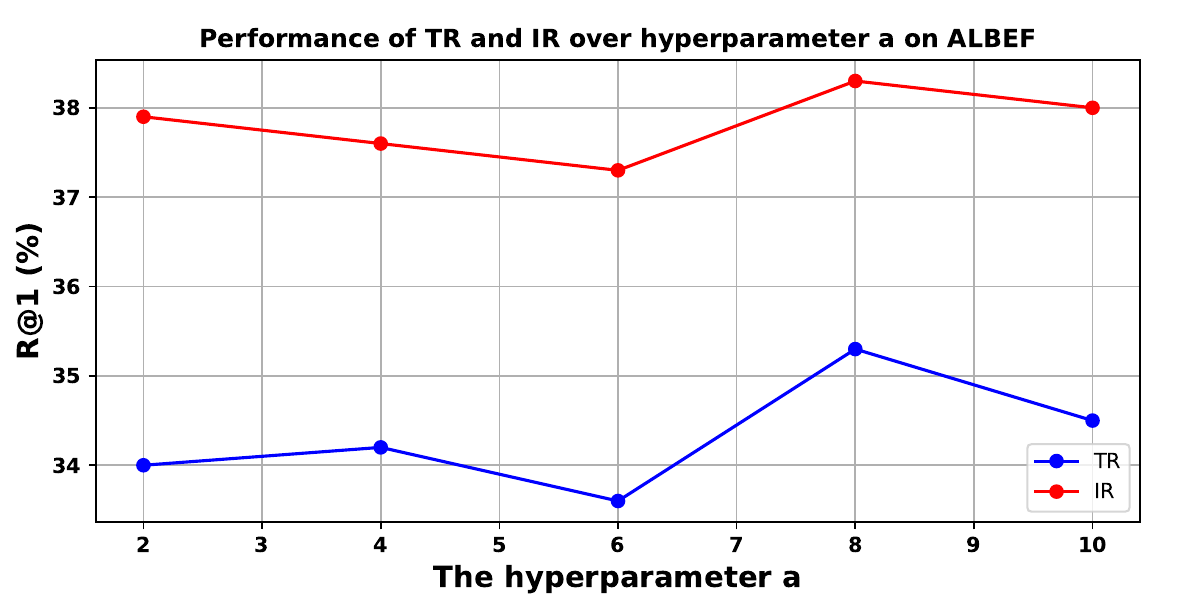}%
\label{fig:ALBEF_response}
}
\subfloat[]{\includegraphics[width=0.45\textwidth]{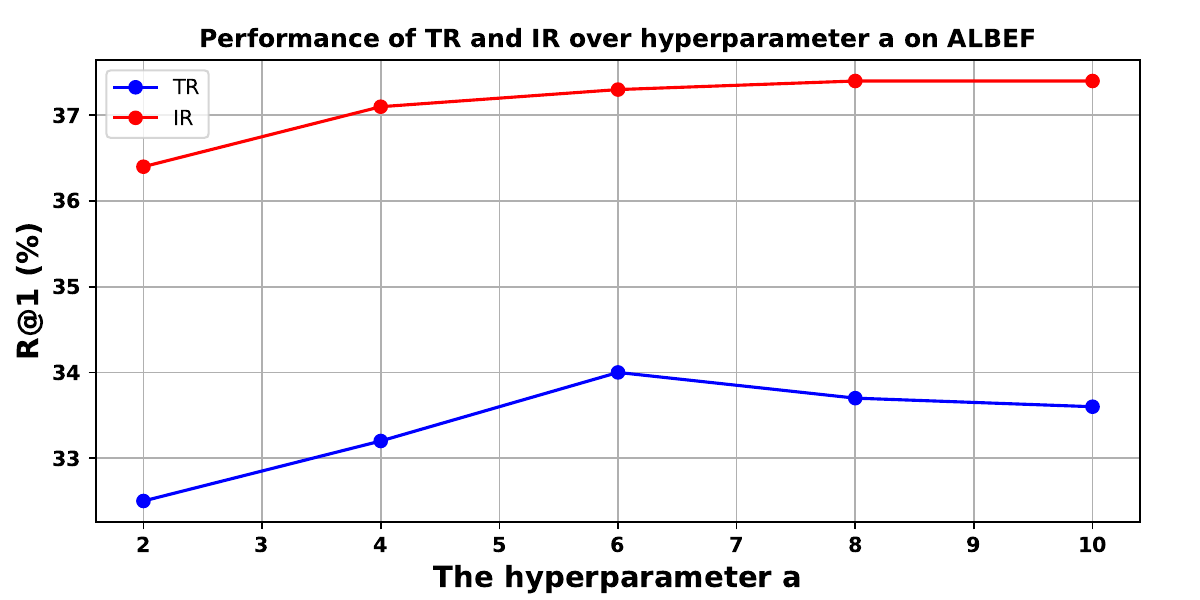}%
\label{fig:TCL_response}
}
\caption{Effect of different hyperparameter $a$. The horizontal coordinate indicates the value of hyperparameter $a$ and the left vertical coordinate indicates the success rate of the attack at R@1 for the image-text retrieval task in Flickr30K dataset.}
\label{fig:hyperparameter_response}
\end{figure*}
We conducted a sensitivity analysis to understand the reason behind the selected values of hyperparameters $a, b, c, d,$ and $g$ in the VLP-attack. 

Initially, all hyperparameters were set to $a = b = c = d = g = 1$. Hyperparameter $a$ is pivotal as it regulates the adversarial loss $L_{adv}$, balancing effective perturbation with perceptibility constraints. We assessed the performance of a transfer-based attack varying the value of $a$, while maintaining $b = c = d = g = 1$, on the Flickr30k dataset for image-text retrieval. The results, depicted in Figure~\ref{fig:hyperparameter_response}, indicate that the transferability of multimodal adversarial samples is optimal at $a = 8$. This setting significantly outperforms other values and surpasses baseline methods, as detailed in Table~\ref{tab: image-text-retrieval}. Consequently, we selected $a = 8$ and $b = c = d = g = 1$ for the image-text retrieval task, which also applies to the visual entailment task.

The detailed analysis on the data augmentation frequency and image transformations in the VLP-attack are provided in Appendix~\uppercase\expandafter{\romannumeral7}, offering deeper insights into the dynamics and effectiveness of the proposed VLP-attack.

\subsection{The Flatness of Loss Landscape Visualization }

\begin{figure}[ht]
\centering\includegraphics[width=3.5in]{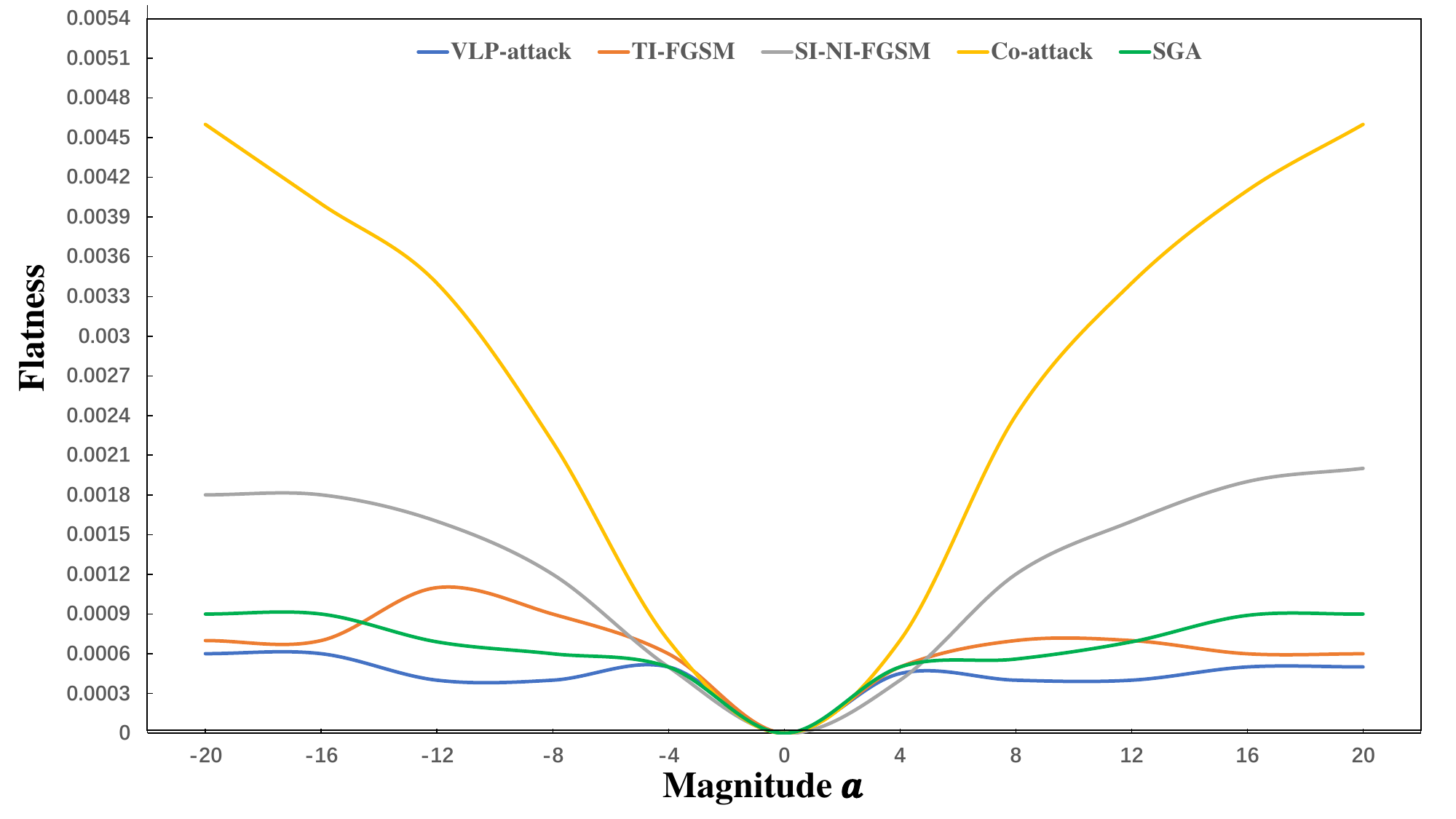}
\caption{
The flatness visualization of adversarial samples in visual entailment task.}
 \label{fig:flatness_visualization}
\end{figure}
We visualize the flatness of the loss landscape~\cite{qin2022boosting} around $x'_i$ on CLIP for VLP-attack(our method), SGA, Co-attack, TI-FGSM, and SI-NI-FGSM by plotting the loss change when moving $x'_i$ along a random direction with various magnitudes in the visual entailment task. Specifically, firstly, we sample $p$ from a Gaussian distribution and normalize $p$ on a $l_2$ unit norm ball, $p \leftarrow{\frac{p}{||p||_F}}$. Then we use different magnitudes $a$ to calculate the loss change(flatness) $f(a)$. The formula is as follows:
\begin{equation}
    f(a) = \mathrm{Sim} (\mathcal{F}_s(x'_i + a \cdot p), \mathcal{F}_s(x'_t) )  - \mathrm{Sim} (\mathcal{F}_s(x'_i), \mathcal{F}_s(x'_t))
\end{equation}
where Sim denotes the function for similarity measurement. Due to the $p$ being randomly sampled, we calculate 20 times with different $p$ and average the value to visualize the flatness of the loss landscape around $x'_i$.

From Figure~\ref{fig:flatness_visualization}, we can observe that compared to the baselines, the adversarial images crafted by VLP-attack are located at a much flatter region, which reflects that our method can discover highly robust adversarial examples that exhibit minimal sensitivity to variations in the decision boundary. This effectively reduces the potential overfitting of the surrogate model and improves the transferability of adversarial samples compared with other baselines.

% \begin{figure}[!ht]
% \centering\includegraphics[width=3.5in]{images/transformation_run_time_analysis_2.pdf}
% \caption{
% Effect of images transformation number. The horizontal coordinate indicates the number of images transformation and the left vertical coordinate indicates the success rate of the attack at R@1 for the image-text retrieval task in Flickr30K dataset. The right vertical coordinate indicates the average runtime of different image transformation numbers in VLP-attack when attacking on every image-text pair. The runtime is in seconds.}
%  \label{fig:images_transoformation_retrieval}
% \end{figure}

\subsection{Visualization}

\begin{figure*}[ht]
\centering\includegraphics[width=5in]{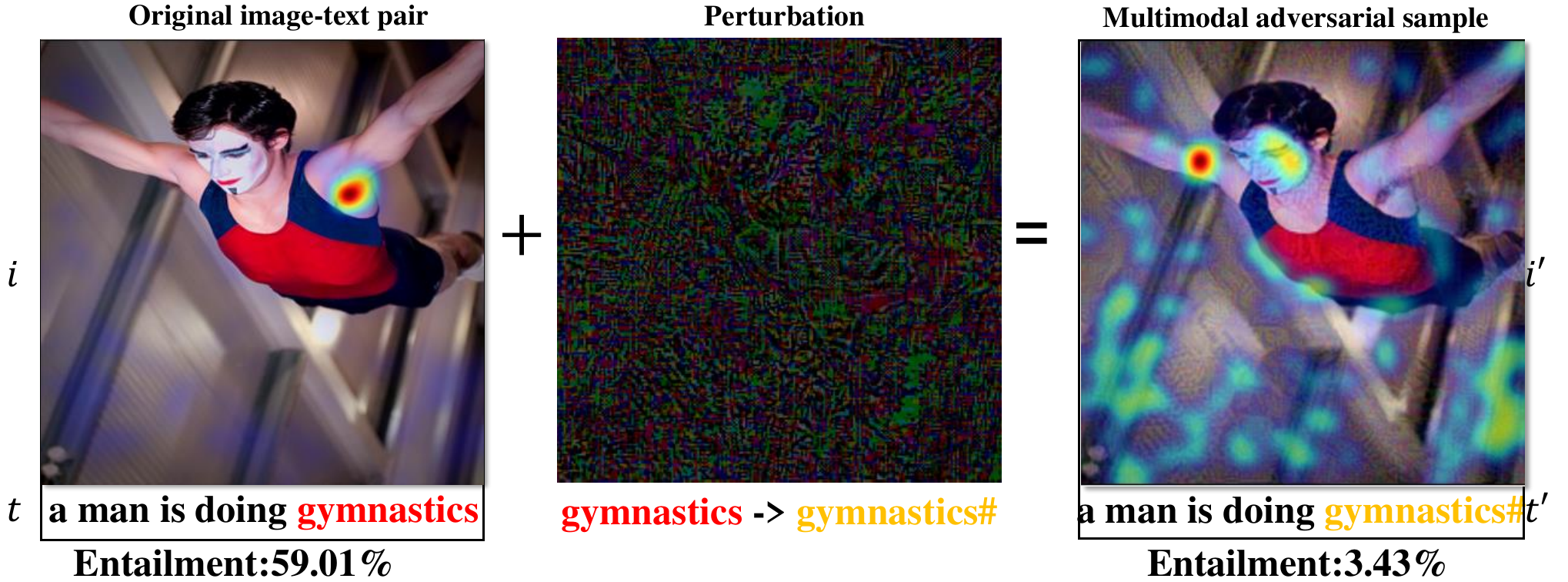}
\caption{The Grad-CAM visualizations of  the original image-text pair, the multimodal adversarial sample crafted by VLP-attack against ALBEF on SNLI-VE dataset for visual entailment task, where the adversarial perturbation is obtained by $x'_i - x_i$ ( pixel values of perturbation are
amplified ×5 for visualization ).}
 \label{fig:visualization_ve}
\end{figure*}

To better understand VLP-attack, we provide the Grad-CAM~\cite{selvaraju2017grad} visualization for the multimodal adversarial samples, 
against ALBEF for the visual entailment task in Figure~\ref{fig:visualization_ve}.
% and the visualizations  for the image-text retrieval task are put in Appendix C. 
The heat map from Grad-CAM can show the attention of the target model for the images when making decisions.

Compared the origin image-text pair with the multimodal adversarial samples in Figure~\ref{fig:visualization_ve} for the visual entailment task, we can observe that the prediction score of ALBEF for the ``entailment'' is reduced from 59.01\% to 3.43\% when the perturbation is added to the original image-text pair and the perturbation is in an imperceptible constraint. 
More visualization examples can be found in Appendix~\uppercase\expandafter{\romannumeral8}.

\subsection{Discussion}
We briefly discuss the importance of the work, why contrastive learning can help to generate multimodal adversarial samples with high transferability and the limitations of our work.

% \textbf{The importance of multimodal adversarial attacks.}
% % We briefly discuss the importance of the work and why contrastive learning can help generate multimodal adversarial samples with high transferability. 
% Historically, research on adversarial attacks has concentrated on single-modal data, with multimodal large models primarily studied within white-box attack scenarios. However, multimodal adversarial attacks pose a more significant threat than single-modal attacks, as they can more profoundly affect the models' ability to interpret and understand visual-language content as shown in Figure~\ref{fig:intr}. As multimodal models play a larger role in the real world, their adversarial robustness is becoming a more pressing issue.

\textbf{The unique challenge of studying the transferability of multimodal adversarial samples.}
Historically, research on adversarial attacks has concentrated on single-modal data, with multimodal large models primarily studied within white-box attack scenarios. However, multimodal adversarial attacks pose a more significant threat than single-modal attacks, as they can more profoundly affect the models' ability to interpret and understand visual-language content as shown in Figure~\ref{fig:intr}. As multimodal models play a larger role in the real world, their adversarial robustness in a black-box setting is becoming a more pressing issue.
Enhancing the transferability of multimodal adversarial samples presents several unique challenges compared to adversarial images alone:
\begin{itemize}
    \item \textbf{Complex Interactions Between Modalities:} In vision-language models, interactions between text, image, audio, or other modalities add layers of complexity. Perturbations need to affect multiple data types simultaneously while maintaining coherence across modalities. Ensuring that adversarial modifications in one modality (e.g., image) effectively influence the model when processed alongside another modality (e.g., text) is significantly challenging.
    \item \textbf{Differing Vulnerabilities Across Modalities:} Each modality (e.g., vision, language) has unique adversarial vulnerabilities and response characteristics. For example, while images may tolerate minor pixel-level changes, text modifications must maintain semantic integrity. Crafting perturbations that work across these differing sensitivities complicates transferability across multimodal models.
    \item \textbf{Model Architecture Variability:} Vision-language models vary widely in architecture, especially in how they fuse and align features from different modalities. Perturbations transferable to one model architecture may not work effectively on another due to differences in how models process and integrate multimodal inputs.
\end{itemize}

\textbf{Strategies for Enhancing the Transferability of Multimodal Adversarial samples in VLP Models.}
As VLP models grow in size and complexity, maintaining the transferability of multimodal adversarial samples from vision-language perturbations becomes increasingly challenging. Here are several strategies to mitigate this issue and enhance the effectiveness of multimodal adversarial samples across diverse VLMs:
\begin{itemize} 
    \item \textbf{Utilize Larger Surrogate VLP Models}: Employing a surrogate model of similar size is the most straightforward and effective method to maintain transferability. Generally, models of similar dimensions exhibit analogous decision boundaries, allowing adversarial samples generated by these surrogates to transfer more effectively to the target model. 
    \item \textbf{Ensemble Adversarial Attacks}: Implementing ensemble methods to generate adversarial samples can enhance their robustness and transferability. By combining adversarial inputs from multiple attack strategies or models, it is possible to create complex adversarial samples that large VLP models may not be equipped to handle.
    \item \textbf{Adaptive Adversarial Fine-tuning}: Fine-tuning a model adversarially across a diverse set of models, including those of varying sizes and architectures, exposes the surrogate model to a broad spectrum of defenses and complexities. This exposure increases the likelihood that the generated adversarial samples will be effective across different vision-language models. 
\end{itemize}

\textbf{The transferability of contrastive learning.}
Contrastive learning is used to encourage the adversarial samples away from the semantics of original samples from different views, which benefits to perturb the essential and general characteristics of the samples rather than overfitting to model-specific features on surrogate models as show in Figure~\ref{fig:flatness_visualization}. This robustness is crucial for transferring adversarial samples to different victim models.

\textbf{Limitations of this work.} Contrastive learning significantly depends on data augmentation to generate positive and negative samples. Ineffectively selected negative samples can induce suboptimal feature perturbations, diminishing the transferability of multimodal adversarial samples. Therefore, the efficacy of positive and negative sample selection is critical in enhancing VLP-attack transferability, presenting a notable challenge in this domain.

\textbf{The potential impact of this work.}
Adversarial attacks serve a crucial role in identifying weaknesses and assessing the robustness of deep learning models. Our research highlights the essentiality of multimodal adversarial attacks in revealing vulnerabilities associated with integrating diverse modalities in VLP mdoels. By systematically exploring these attacks, researchers can pinpoint specific weaknesses and develop models designed to withstand such adversarial tactics. Moreover, as VLP models are increasingly utilized in real-world applications, enhancing their resilience against adversarial threats is imperative to maintain public trust and acceptance. This is particularly vital in critical areas such as image-text retrieval and misinformation detection, where the reliability of VLMs is essential to prevent biased or detrimental outcomes.

\section{Conclusion}
\label{sec:conclusion}
This paper first proposes a gradient-based multimodal adversarial attack against VLP models in a black-box setting.
Compared with the single-modal adversarial attack methods, our method can generate both adversarial
text and adversarial images simultaneously, and leverages image-text contrastive loss and intra-modal contrastive loss to improve the transferability of multimodal adversarial samples. 
The experimental results show that our method outperforms other baselines in image-text retrieval tasks and visual entailment tasks.

\bibliographystyle{IEEEtran}
\bibliography{egbib.bib}

\begin{IEEEbiography}[{\includegraphics[width=1in,height=1.25in,clip,keepaspectratio]{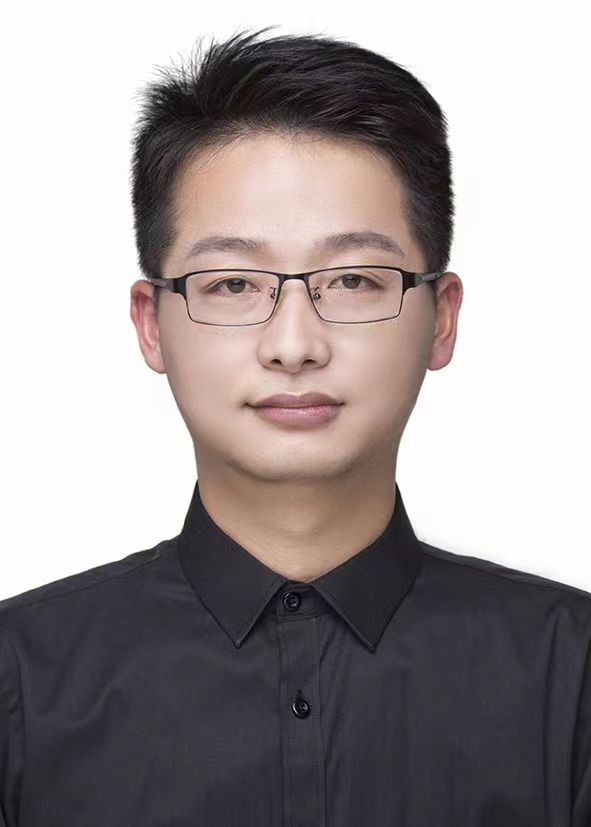}}]{Youze Wang}
received a B.S. and master's degree from the School of Computer Science and Information Engineering at Hefei University of Technology, Hefei, China, where he is currently working toward his Ph.D. degree. His research interests include multimodal computing and multimodal adversarial robustness in machine learning. 
\end{IEEEbiography}
\vspace{-1cm}

\begin{IEEEbiography}[{\includegraphics[width=1in,height=1.25in, clip,keepaspectratio]{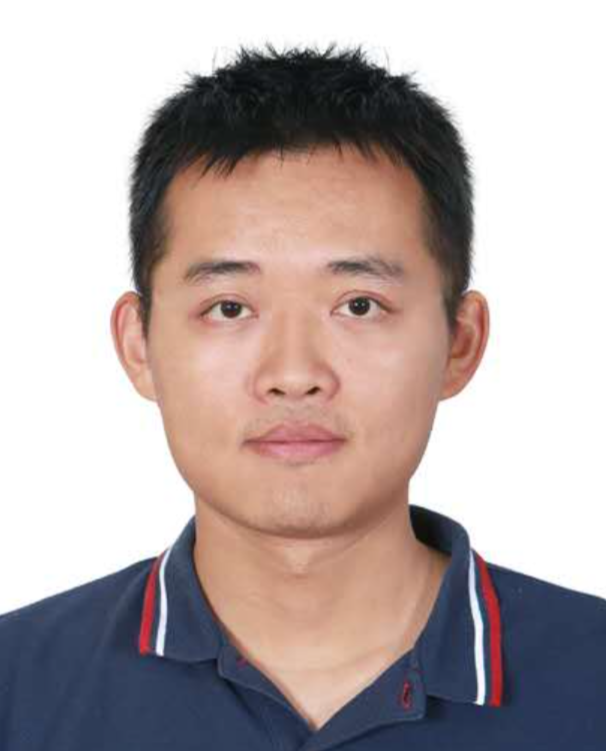}}]{Wenbo Hu} is an associate professor in Hefei University of Technology. He received a Ph.D. degree from Tsinghua University in 2018. His research interests lie in machine learning, especially probabilistic machine learning and uncertainty, generative AI, and AI security. He has published more than 20 peer-reviewed papers in prestigious conferences and journals, including NeurIPS, KDD, IJCAI, etc.
\end{IEEEbiography}
\vspace{-1cm}

\begin{IEEEbiography}[{\includegraphics[width=1in,height=1.15in, clip,keepaspectratio]{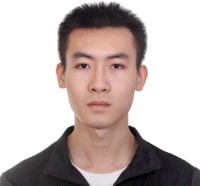}}]{Yinpeng Dong} received the B.S. and Ph.D. degrees from the Department of Computer Science and Technology, Tsinghua University. He is currently a Post-Doctoral Researcher with the Department of
Computer Science and Technology, Tsinghua University. His research interests include the adversarial robustness of machine learning and deep learning. He received the Microsoft Research Asia Fellowship
and the Baidu Fellowship.
\end{IEEEbiography}
\vspace{-1cm}

\begin{IEEEbiography}[{\includegraphics[width=1in,height=1.15in, clip,keepaspectratio]{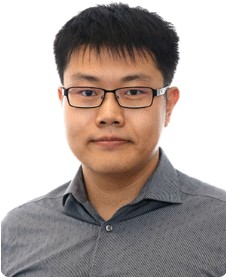}}]{Hanwang Zhang} received the BEng.(Hons.) degree in computer science from Zhejiang University, Hangzhou, China, in 2009, and a Ph.D. degree in computer science from the National University of
Singapore (NUS), Singapore, in 2014. He is currently an associate professor with Nanyang Technological University, Singapore. His research interests include developing multi-media and computer vision techniques for efficient search and recognition of visual content. He received the Best Demo Runner-Up Award in ACM MM 2012 and the Best Student Paper Award in ACM MM 2013. He was the recipient of the Best Ph.D. Thesis Award of the School of Computing, NUS, 2014.
\end{IEEEbiography}
\vspace{-1cm}

\begin{IEEEbiography}[{\includegraphics[width=1in,height=1.15in, clip,keepaspectratio]{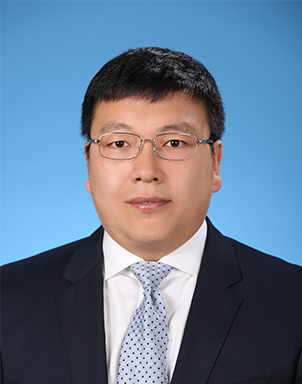}}]{Hang Su} is an associated professor in the department of computer science and technology at Tsinghua University. His research interests lie in the adversarial machine learning and robust computer vision, based on which he has published more than 50 papers including CVPR, ECCV, TMI, etc. He has served as area chair in NeurIPS and the workshop co-chair in AAAI22. He received “Young Investigator Award” from MICCAI2012, the “Best Paper Award” in AVSS2012, and “Platinum Best Paper Award” in ICME2018.
\end{IEEEbiography}
\vspace{-1cm}

\begin{IEEEbiography}[{\includegraphics[width=1in,height=1.25in,clip,keepaspectratio]{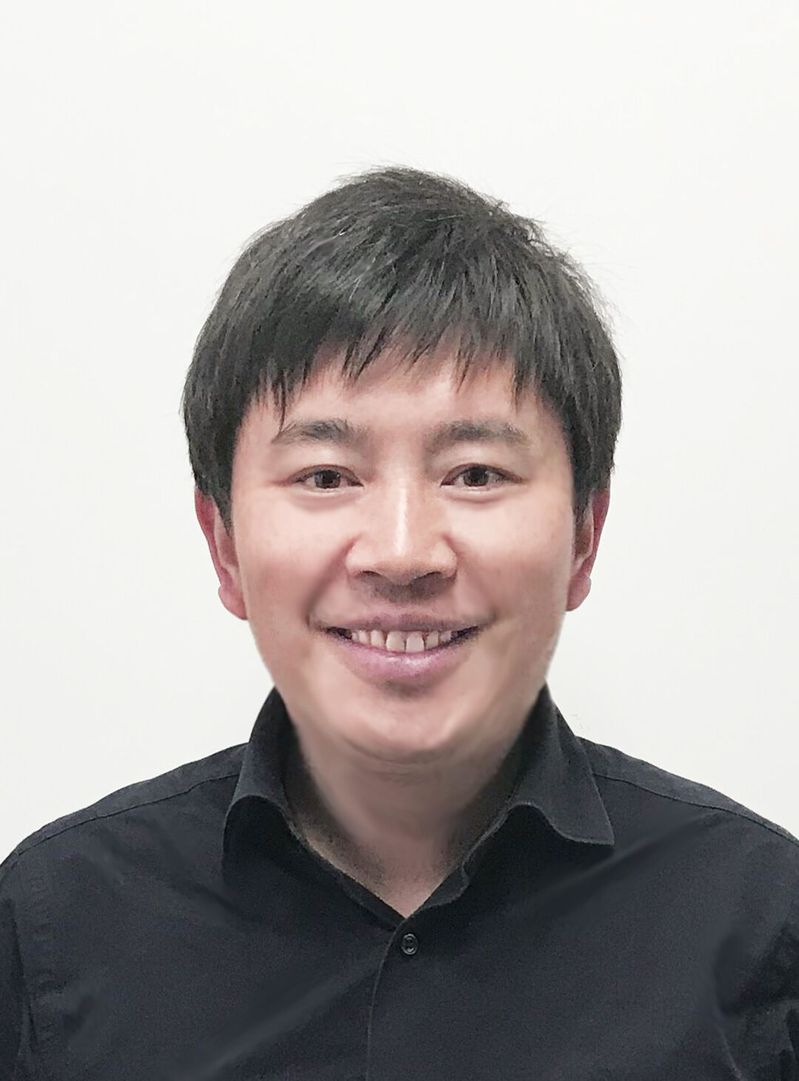}}]{Richang Hong (Member, IEEE)}
received a Ph.D. degree from the University of Science and Technology of China, Hefei, China, in 2008. He was a Research Fellow of the School of Computing at the National University of Singapore, from 2008 to 2010. He is currently a Professor at the Hefei University of Technology, Hefei. He is also with the Key Laboratory of Knowledge Engineering with Big Data (Hefei University of Technology), Ministry of Education. He has coauthored over 100 publications in the areas of his research interests, which include multimedia content analysis and social media. He is a member of the ACM and the Executive Committee Member of the ACM SIGMM China Chapter. He was a recipient of the Best Paper Award from the ACM Multimedia 2010, the Best Paper Award from the ACM ICMR 2015, and the Honorable Mention of the IEEE Transactions on Multimedia Best Paper Award. He has served as the Technical Program Chair of the MMM 2016, ICIMCS 2017, and PCM 2018. Currently, he is an Associate Editor of IEEE Transactions on Big Data, IEEE Transactions on Computational Social Systems, ACM Transactions on Multimedia Computing Communications and Applications, Information Sciences (Elsevier), Neural Processing Letter (Springer) and Signal Processing (Elsevier).
\end{IEEEbiography}
\clearpage

\twocolumn[
\begin{@twocolumnfalse}
\section*{\centering{ Supplementary Material
 :\\ Exploring Transferability of Multimodal Adversarial Samples for Vision-Language Pre-training Models with Contrastive Learning\\[25pt]}}
\end{@twocolumnfalse}
]

\section{The pseudocode of VLP-attack algorithm}
\label{sec:appendix_algorithm}
VLP-attack optimizes the perturbation and generates the multimodal adversarial samples simultaneously by unifying them into a gradient-based framework, which benefits to destroy the alignment of image-text pairs. In addition, we use a contrastive learning mechanism to perturb the inherent structures in the benign samples and the contextual integrity of image-text pairs from various views to improve the transferability of multimodal adversarial samples, all of which is outlined in Algorithm~\ref{alg: algorithm}.
% ------------------------------------------------------------------------------------------
\begin{algorithm}[ht]
	\renewcommand{\algorithmicrequire}{\textbf{Input:}}
	\renewcommand{\algorithmicensure}{\textbf{Output:}}
	\caption{Transfer-based Multimodal Adversarial Attack}
	\label{alg: algorithm}
	\begin{algorithmic}[1]
		\REQUIRE 
        A surrogate model $\mathcal{F}_s$; a benign sample $X = \{x_i, x_t\}$, and ground-truth label $y$; the size of perturbation $\epsilon$; learning rate $r$; iterations $H$; the number of images transformation $M$; the probability $p$ of images transformation;
		\ENSURE  A multimodal adversarial sample $X' = \{x'_i, x'_t\}$;
		\STATE Initialize the matrix $\Theta$ by $x_t$, which is used to sample adversarial texts; 
         \STATE $\alpha _i = \epsilon/H$; $\alpha _\Theta = r$; $x'_i = x_i$; $m_{\Theta} = \gamma$;
         \STATE Generate the images augmentation and texts augmentation;
        \FOR { $t$ = 0 in $H$ - 1}
        \STATE $g_{x'_i}$ = 0
        \FOR {$k=0$ in $M$}
        \STATE  Input ($x_i, x_t$) to $\mathcal{F}_s$ and calculate the total loss $L$ according to eq (9);
        \STATE Obtain the gradient of $x'_i$: $\nabla _{x'_i} L(\mathcal{F}_s(T(x'_i, p)), x'_t)$;
        \STATE Obtain the gradient of $\Theta$: $\nabla _{\Theta} L(\mathcal{F}_s(T(x'_i, p)), x'_t)$;
        \STATE Sum the gradients as $g_{x'_i} = g_{x'_i} + \nabla _{x'_i} L(\mathcal{F}_s(T(x'_i, p)), x'_t)$
        \STATE Update the weights of matrix $\Theta$ according to eq (10);
        \ENDFOR
            \STATE Average the gradients of $x'_i$: $g_{x'_i} = \frac{1}{M} \cdot g_{x'_i}$ ;
        \
    \STATE Update the perturbation $\delta$ on $x_i$ according to eq (11);
        \ENDFOR
        \STATE  Sampling an  $\pi$ from the distribution $P_{\Theta}$;
        \STATE Obtain the tokens ids in the adversarial text $x'_t$: argmax$(\pi)$;
        \STATE Obtain the adversarial image $x'_i$: $x'_i = x_i + \delta$;
        \RETURN $(x'_i, x'_t)$;
    \end{algorithmic}
\end{algorithm}

% ------------------------------------------------------------------------------------------------------

\section{Sensitivity Analysis }
\label{sec: appendix_sensitivity_analysis}
We conducted a comprehensive sensitivity analysis to elucidate the influence of data augmentation number and the number of image transformations on the VLP-attack. 

\subsubsection{Impact of Data Augmentation Number}
\begin{figure}[ht]
\centering\includegraphics[width=3.5in]{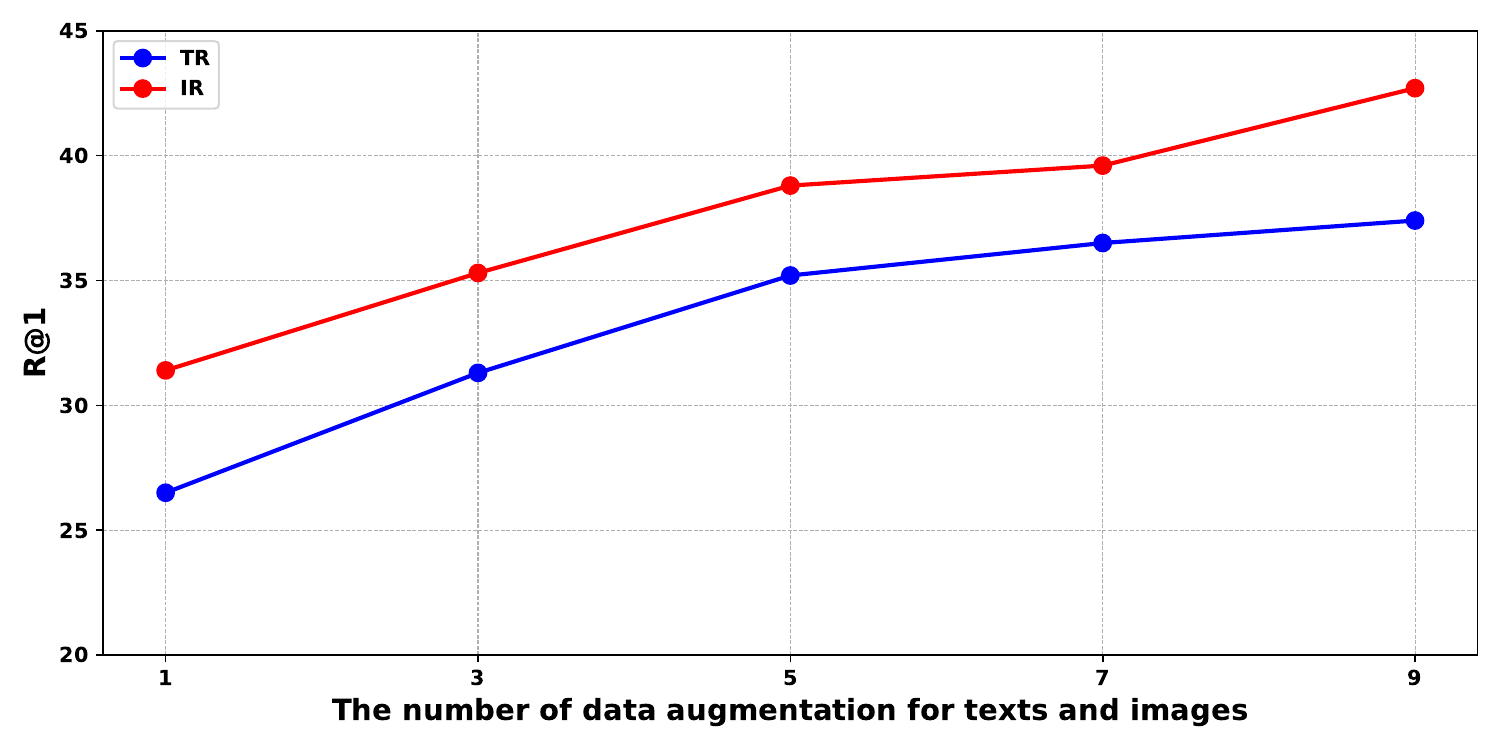}
\caption{Effect of data augmentation number. The horizontal coordinate indicates the number of data augmentation and the vertical coordinate indicates the success rate of the attack at R@1 for the image-text retrieval task in Flickr30K dataset.}
 \label{fig:data_augmentation_retrieval}
\end{figure}

To verify the effect of different data augmentation numbers in VLP-attack, we vary the number of data augmentation from 1 to 9 to craft the multimodal adversarial samples and attack ALBEF on Flickr30k dataset for the image-text retrieval task. From Figure~\ref{fig:data_augmentation_retrieval}, we can observe that the attack success rate in R@1 of TR and IR increases with more data augmentation. However, the rate of increase in attack success rates is gradually slowing down, suggesting that on the one hand, adding more augmentation data does not lead to significant performance gains, and on the other hand, VLP-attack may require more fine-grained augmentation data. To balance the cost time and attack success rate, we set data augmentation numbers = 7 in VLP-attack.

\subsubsection{Impact of Images Transformation Number}

\begin{figure}[!ht]
\centering\includegraphics[width=3.5in]{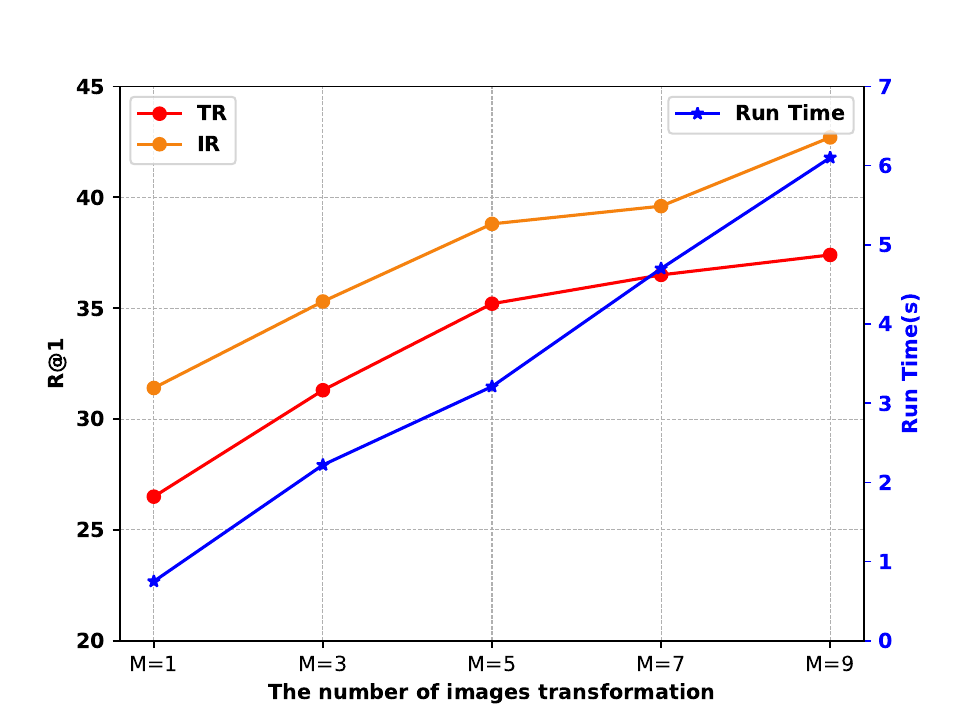}
\caption{
Effect of images transformation number. The horizontal coordinate indicates the number of images transformation and the left vertical coordinate indicates the success rate of the attack at R@1 for the image-text retrieval task in Flickr30K dataset. The right vertical coordinate indicates the average runtime of different image transformation numbers in VLP-attack when attacking on every image-text pair. The runtime is in seconds.}
 \label{fig:images_transoformation_retrieval}
\end{figure}

We pad the images with a given probability as the diverse input to improve the transferability of adversarial images. To analyze the effect of image transformation numbers in VLP-attack, we vary the number of images transformation from 1 to 9 to generate the multimodal adversarial samples in the test set of the Flickr30k dataset. The results of attacking ALBEF using the above adversarial samples are shown in Figure~\ref{fig:images_transoformation_retrieval}. We can observe that R@1 in TR and IR increases with more image transformation numbers, and runtime is gradually increasing due to more image transformations. To balance the cost time and attack success rate, we set image transformation numbers = 5 in VLP-attack.

The above analysis reveals that for image-text retrieval tasks, the Attack Success Rate for both Text Retrieval and Image Retrieval improves with an increase in image transformations. However, this leads to higher computational times. 
Furthermore, our method's time complexity is expressed as  \textit{O(H*M*n)}, with \textit{H} as the number of main iterations, \textit{M} as the number of image transformations, and space complexity as \textit{O(n)}, providing insight into the computational resource requirements.

\section{Visualization }
\label{sec:visualization_appendix}
\begin{figure*}[ht]
\centering\includegraphics[width=5in]{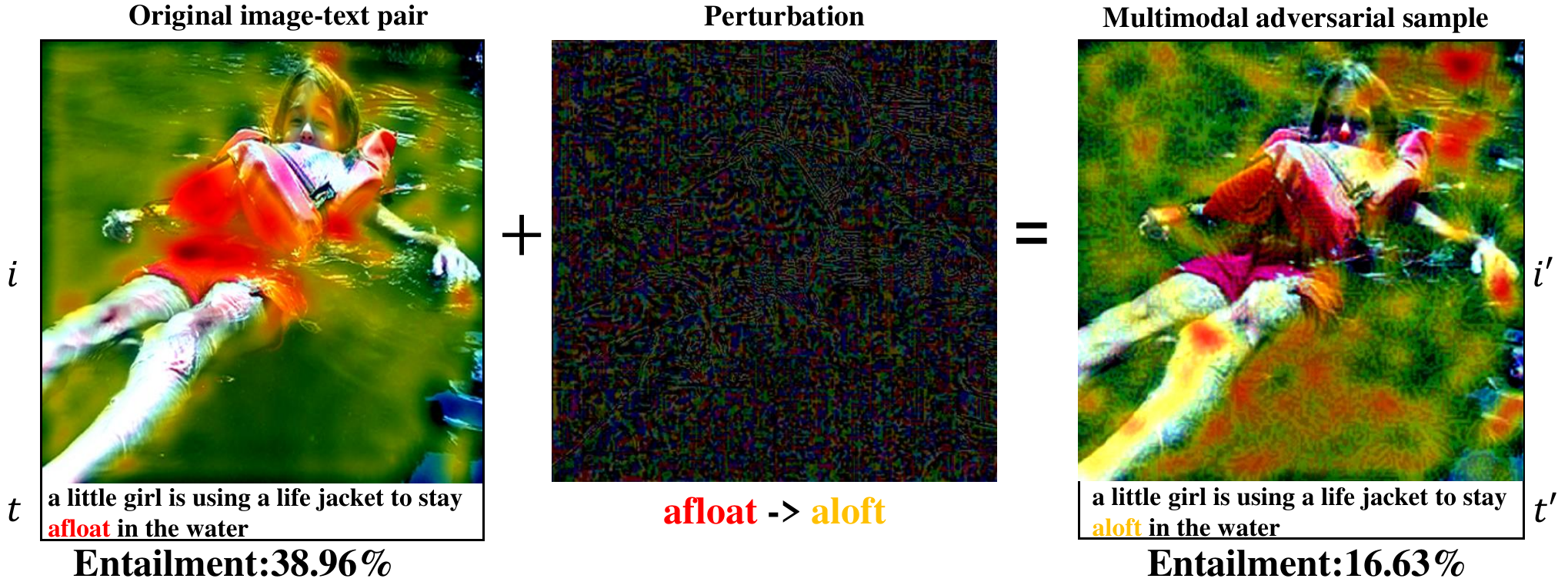}
\caption{The Grad-CAM visualizations of  the original image-text pair, the multimodal adversarial sample crafted by VLP-attack against ALBEF on SNLI-VE dataset for visual entailment task, where the adversarial perturbation is obtained by $x'_i - x_i$ ( pixel values of perturbation are
amplified ×5 for visualization ).}
 \label{fig:visualization_ve-response-1}
\end{figure*}

\begin{figure*}[ht]
\centering\includegraphics[width=5in]{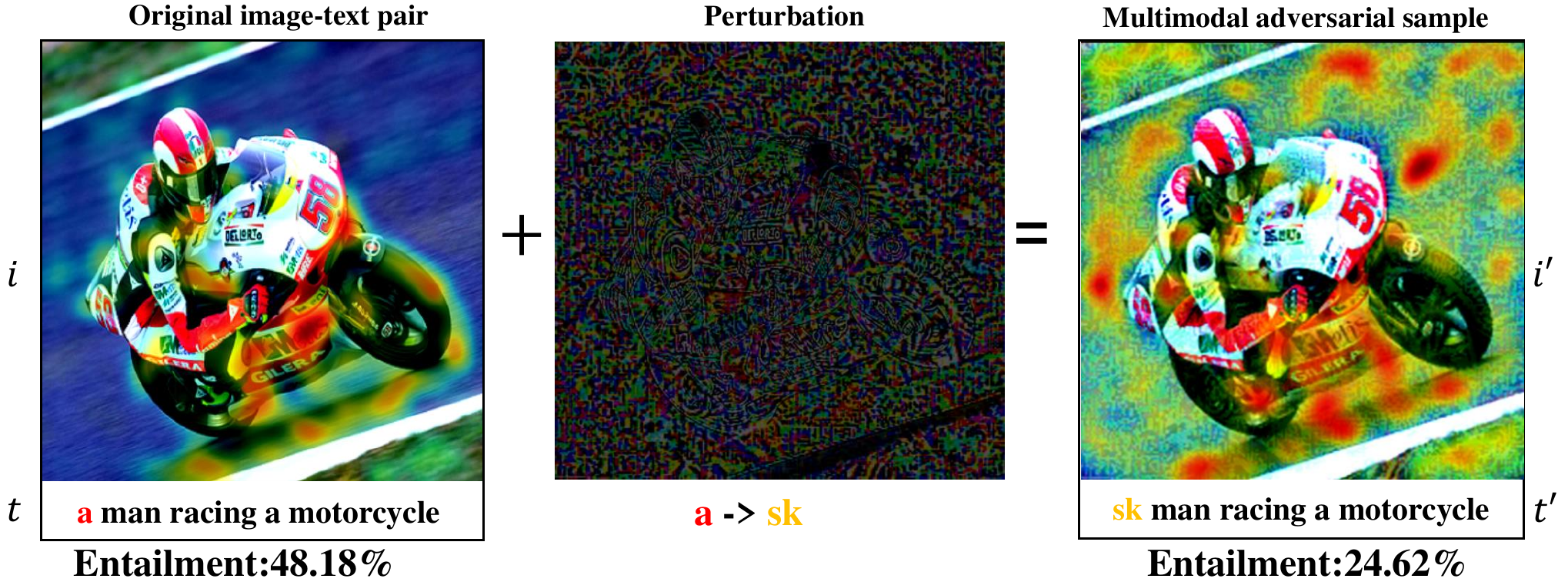}
\caption{The Grad-CAM visualizations of  the original image-text pair, the multimodal adversarial sample crafted by VLP-attack against TCL on SNLI-VE dataset for visual entailment task, where the adversarial perturbation is obtained by $x'_i - x_i$ ( pixel values of perturbation are
amplified ×5 for visualization ).}
 \label{fig:visualization_ve-response-2}
\end{figure*}

To better understand VLP-attack, we provide more Grad-CAM~\cite{selvaraju2017grad} visualization for the multimodal adversarial samples, 
against ALBEF and TCL for the visual entailment task in Figure~\ref{fig:visualization_ve-response-1} and Figure~\ref{fig:visualization_ve-response-2}.
The heat map from Grad-CAM can show the attention of the target model for the images when making decisions.

\vfill

% \fi

\end{document}